\documentclass[11pt,a4paper]{article}

\usepackage[margin=2.5cm]{geometry}
\usepackage[T1]{fontenc}
\usepackage[utf8]{inputenc}
\usepackage{lmodern}
\usepackage{amsmath,amssymb}
\usepackage{graphicx}
\usepackage{subcaption}
\usepackage[table]{xcolor}
\usepackage{booktabs,tabularx}
\usepackage{multirow}
\usepackage{scalefnt}
\usepackage[authoryear,round]{natbib}
\usepackage{hyperref}
\usepackage{authblk}

\hypersetup{
    colorlinks=true,
    linkcolor=blue,
    citecolor=blue,
    urlcolor=blue
}

\title{Anomaly detection in European cryptocurrency exchange-traded products}

\author[1,2]{Julia Kończal\thanks{Corresponding author: julia.konczal@pwr.edu.pl; ORCID: 0009-0009-7237-948X}}
\author[1,2]{Rafał Połoczański\thanks{ORCID: 0000-0003-4214-5892}}

\affil[1]{Faculty of Pure and Applied Mathematics, Hugo Steinhaus Center, Wrocław University of Science and Technology, Wrocław, Poland}
\affil[2]{xyt, Smart Data. Intelligent Trading, Wrocław, Poland}

\date{}

\begin{document}

\maketitle

\begin{abstract}
Cryptocurrency exchange-traded products (ETPs) listed on European exchanges provide a regulated environment for studying intraday market anomalies. We study four Bitcoin and Ethereum ETPs traded on Xetra and Nasdaq Stockholm over the period January 2024 -- December 2025 using one-minute bars. As a benchmark, we adopt an extreme value theory approach in which anomalous bars are defined as returns falling below a threshold estimated by fitting a generalised Pareto distribution to left-tail exceedances. We then propose three new binary anomaly indicators. The first, a cross-venue divergence anomaly, identifies venue-specific price divergence between the two exchanges.
The second is a no-recovery anomaly that identifies extreme price drops followed by little or no recovery over the next ten active bars.
The third is a momentum-reversal anomaly that identifies extreme price drops following positive short-term momentum.
Although each anomaly type represents fewer than 1\% of one-minute bars, statistical analysis using Mann--Whitney U tests shows that anomaly observations exhibit significantly higher effective spreads, higher values of liquidity-related ratios, and more pronounced order-flow imbalances than non-anomalous bars. Furthermore, employing an out-of-sample prediction methodology with four classifiers -- random forest, logistic regression, extreme gradient boosting, and light gradient boosting machine -- shows that all four anomaly types are predictable one bar ahead, with AUC--ROC values of up to 0.82. Permutation importance indicates that short-term volatility and drawdown measures are generally more useful for prediction than microstructure variables.
\end{abstract}

\noindent\textbf{Keywords:} cryptocurrency; extreme value theory; anomaly detection; forecasting; intraday patterns

\vspace{1em}
\section{Introduction}
\label{sec:intro}

Over the past decade, the growth of cryptocurrency markets has led to the development of exchange-traded products (ETPs) that track digital assets.
 Unlike many spot cryptocurrency markets, which trade continuously and often operate outside traditional exchange structures, European cryptocurrency ETPs are listed on regulated venues and trade during standard exchange hours. They allow retail and institutional investors to gain exposure to Bitcoin and Ethereum through regular brokerage accounts \citep{madhavan2014exchange, orlando2025exchange}.
 By 2024, more than 200 cryptocurrency ETPs were listed on major European exchanges, including Xetra, Nasdaq Stockholm, SIX Swiss Exchange, and Euronext Paris. This makes Europe one of the largest regulated markets for cryptocurrency investment products.

The literature on cryptocurrency markets is extensive and addresses various topics, such as return distributions and tail risk \citep{gkillas2018application, konczal2024tail}, market efficiency and return predictability \citep{liu2021risks}, volatility dynamics and clustering \citep{katsiampa2019empirical, song2019cluster}, and cross-market linkages \citep{adelopo2025interconnectedness,kostika2020dynamic}. Studies show that cryptocurrency returns are not normally distributed. The behaviour of cryptocurrency returns differs from the assumptions underlying conventional risk models, particularly because of their fat-tailed distributions, high kurtosis, and persistent volatility \citep{osterrieder2017statistical}.

However, most studies examine spot cryptocurrency markets and use daily data, while regulated ETP markets have received much less attention.
  The few studies that have examined cryptocurrency microstructure at intraday frequencies have focused on bid-ask spreads, order flow, and price discovery \citep{baur2021volatility, easley2026microstructure}, or on the distributional properties of high-frequency returns \citep{foroni2025quantile}. The specific question of how anomalous price behaviour appears in regulated ETP markets, and whether it is detectable and predictable using microstructure information, to our knowledge, has not yet been examined.

Market anomalies have been widely studied in empirical finance.
 In equity markets, momentum \citep{jegadeesh1993returns}, day-of-the-week effects \citep{harris1986transaction}, and flash crashes \citep{kirilenko2017flash} are well-documented examples of market anomalies.
 Cryptocurrency markets exhibit pronounced anomalies, attributed to their high volatility, relatively low liquidity, fragmented trading infrastructure, and the continuous operation of underlying spot markets during exchange hours \citep{hu2019cryptocurrencies}. Calendar effects, weekend anomalies, and momentum reversals have all been documented in daily cryptocurrency data \citep{aharon2019bitcoin, caporale2019day}.

To study these effects, we first need to define which observations should be treated as anomalies. As a benchmark, we use extreme value theory (EVT), specifically the Peaks-over-Threshold (POT) method, which fits a generalised Pareto distribution (GPD) to tail exceedances and has been widely applied to financial returns \citep{doi:10.1177/1471082X241266729, davison1990models, balkema1974residual, pickands1975statistical}. However, the POT approach identifies observations in the far left tail of the return distribution without exploiting the cross-sectional or temporal structure of the data. Since our high-frequency data cover multiple exchanges simultaneously, we can use this information to construct additional anomaly indicators.

Once anomalies are defined, the next question is whether they can be predicted one bar ahead. Machine learning classifiers have been applied to a wide range of financial prediction problems, including fraud detection in high-frequency markets \citep{poutre2024deep} and market stress prediction \citep{li2025machine}. Ensemble methods such as random forests \citep{breiman2001random} and gradient boosting \citep{ramraj2016experimenting} have shown solid performance on financial data \citep{wang2022corporate, ramraj2016experimenting}. In cryptocurrency markets, machine learning has been used to predict high-frequency returns, model order flow, and analyse market microstructure  \citep{li2025machine, foroni2025quantile}.

In this paper we make three contributions. First, we provide a characterisation of the European cryptocurrency ETP market and select four instruments for analysis based on data completeness: VanEck Bitcoin ETP (VBTC.XE) and VanEck Ethereum ETP (VETH.XE), traded on Xetra, and Virtune Bitcoin ETP (VIRBTC.ST) and Virtune Ethereum ETP (VIRSETHS.ST), traded on Nasdaq Stockholm. Second, we propose three new binary anomaly indicators -- cross-venue divergence ($D^{CV}$), no recovery ($D^{NR}$) and momentum reversal ($D^{MR}$) -- which complement the baseline POT definition. Third, we evaluate the one-bar-ahead predictability of all four anomaly classes in an out-of-sample framework with four classifiers -- logistic regression (LR), random forest (RF), extreme gradient boosting (XGBoost), and light gradient boosting machine (LightGBM).

This paper is structured as follows. In Section~\ref{sec:market} we provide an overview of the European cryptocurrency ETP market. Section~\ref{sec:data} describes the data. In Section~\ref{sec:anomaly} we introduce the four anomaly definitions and analyse their statistical properties and microstructure correlates. Section~\ref{sec:prediction} presents the prediction framework and results. Finally, Section~\ref{sec:conclusions} concludes the paper.

  %%% ----------------------------------------
\section{Overview of the European crypto ETP market}
\label{sec:market}

Exchange-traded products include several types of products, such as exchange-traded funds (ETFs), exchange-traded notes (ETNs), and exchange-traded commodities (ETCs). These types differ in their legal structure, main features, and the way they give investors access to underlying assets.
At the broadest level, an ETP refers to an exchange-listed instrument whose value is linked to an underlying asset, index, or strategy \citep{madhavan2014exchange}. The legal form of an ETP affects how investors gain exposure to the underlying assets and what risks they face. In some cases, this exposure is provided through an investment fund. ETFs use this structure and hold a portfolio of underlying assets for their investors. In Europe, ETFs need to follow the Undertakings for Collective Investment in Transferable Securities (UCITS) framework, which sets rules for diversification and investor protection \citep{ESMA2014}.

Other ETPs give investors exposure through debt-based structures instead of ownership in a fund. ETNs are debt securities issued by financial institutions, and their returns depend on the performance of a specific underlying asset or index
\citep{orlando2025exchange}. Investors do not own the underlying asset directly. Instead, they have a claim against the issuer. As a result, they face issuer credit risk as well as market risk. ETCs have a similar structure and are used to provide exposure to commodities. Many ETCs are physically backed. The issuer holds this asset to protect investors if it cannot repay the notes.
This distinction is especially important in the European crypto market. The UCITS framework does not allow an ETF to invest only in a single cryptocurrency. Hence, exchange-listed crypto products in Europe are mainly structured as ETNs or ETCs.

The empirical analysis is based on data obtained from xyt. The initial dataset includes 200 cryptocurrency ETPs listed on European exchanges and covers the period from January 2024 to December 2025. Before we move directly to the selection of instruments for anomaly detection analysis, we first examine the market landscape. This preliminary market overview allows us to describe the structure of the European crypto ETP market. 
The initial dataset includes products from several well-known European crypto ETP providers, such as 21Shares, WisdomTree, ETC Group, CoinShares, Valour, and Virtune.
 None of the instruments in the initial universe qualifies as a UCITS ETF. For this reason, we use the term ETP throughout the study to refer to all instruments in the dataset, while we recognise that most of them are legally structured as ETNs or ETCs.

\begin{figure}[h]
\center\includegraphics[width=0.7\linewidth]{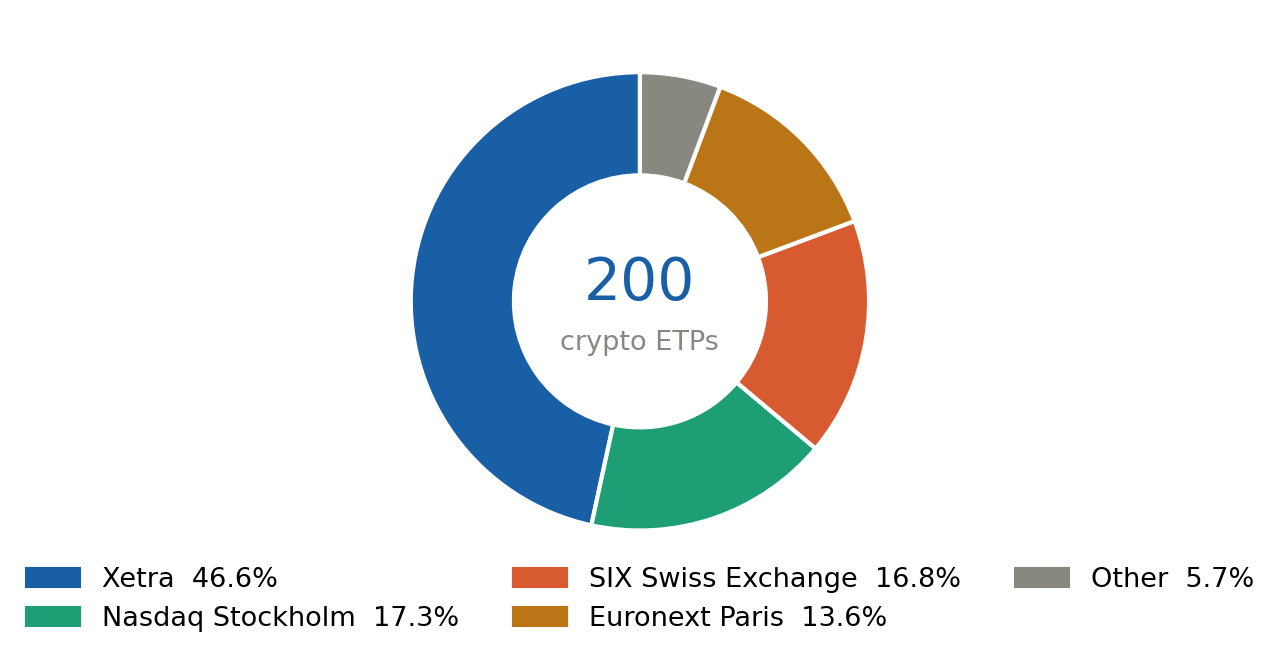}
        \caption{The 200 cryptocurrency ETPs grouped by their main trading venue. Each instrument is assigned to the venue where it had the highest trading volume during the sample period.
}
\label{fig:venue_donut}
\end{figure}

In Figure~\ref{fig:venue_donut} we present the distribution of the 200 cryptocurrency ETPs by primary trading venue. Most crypto ETPs are concentrated on a few major exchanges.
 Xetra is the main trading venue for 46.6\% of the instruments, which makes it the largest venue in the sample.
It is followed by Nasdaq Stockholm, with 17.3\%, SIX Swiss Exchange, with 16.8\%, and Euronext Paris, with 13.6\%. The strong position of Xetra is associated with the presence of German issuers, many of whose products are primarily traded in Frankfurt. On the other hand, Nasdaq Stockholm is more specialised segment of the market. The instruments listed there are almost exclusively Swedish ETPs issued by Valour and Virtune.

\begin{figure}[h]
\center\includegraphics[width=0.7\linewidth]{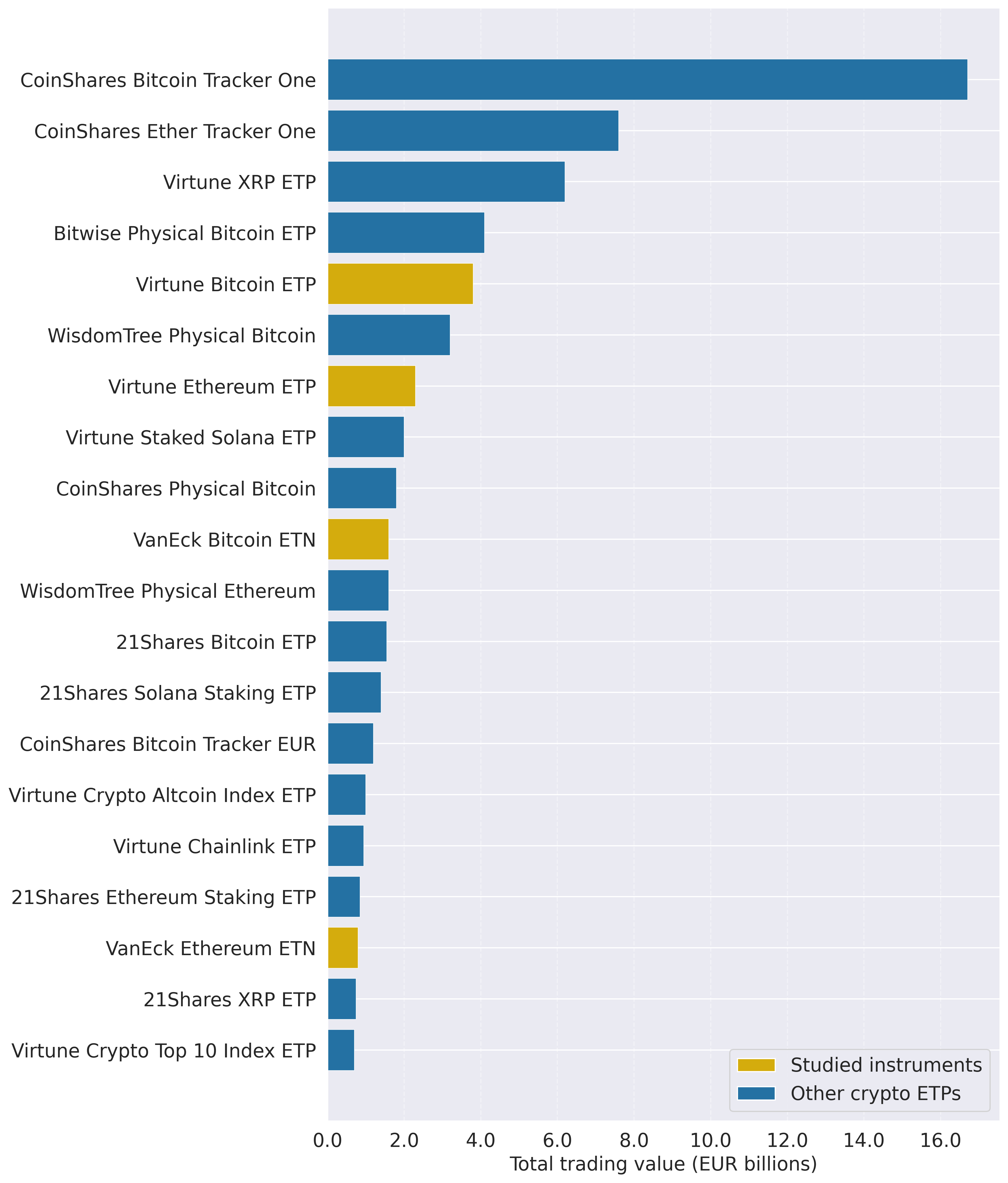}
       \caption{Top 20 cryptocurrency ETPs by trading value over the full sample period. The instruments selected for further analysis are shown in yellow.
 }
\label{fig:top_etps_value}
\end{figure}

In Figure~\ref{fig:top_etps_value} we illustrate the ranking of the 20 most actively traded cryptocurrency ETPs by trading value over the sample period. Trading activity is concentrated at the top of the distribution. The two largest instruments are CoinShares Bitcoin Tracker One and CoinShares Ethereum Tracker One, traded on Nasdaq Stockholm, with trading values of approximately EUR 16.5 billion and EUR 7.5 billion. Beyond these two instruments, values decline sharply.
Overall, these results show that Europe has many listed crypto ETPs, but most trading takes place in only a small number of products.

\section{Data}
\label{sec:data}

 From the initial 
landscape of 200 cryptocurrency ETPs described in Section~\ref{sec:market}, 
we select four instruments for the anomaly detection analysis based on data completeness. We select instruments containing the largest number 
of one-minute bars with non-zero trading activity over the sample period.
The first two instruments are issued by VanEck and traded on Xetra. VanEck 
Bitcoin ETP (VBTC.XE) provides direct exposure to Bitcoin and is one of the 
most established cryptocurrency ETPs available on the German exchange. VanEck 
Ethereum ETP (VETH.XE) tracks the price of Ethereum and follows the same structure.
The remaining two instruments are issued by Virtune and traded on Nasdaq Stockholm. Virtune Bitcoin ETP (VIRBTC.ST) provides exposure to 
Bitcoin and is among the most actively traded cryptocurrency ETPs on the 
Swedish market, as confirmed by Figure~\ref{fig:top_etps_value}. Virtune 
Ethereum ETP (VIRSETHS.ST) tracks the price of Ethereum.

After removing one-minute bars with no trading activity, the final sample contains 155{,}510 observations for VBTC.XE, 99{,}409 for VETH.XE, 115{,}679 for VIRBTC.ST, and 73{,}448 for VIRSETHS.ST.
 The Ethereum instruments have more inactive bars because they are traded less frequently than the Bitcoin products. For each bar, the dataset contains open, high, low, and close prices, total trading volume, trading value in EUR, bid and ask volumes, dark pool volume, periodic auction volume, and the effective spread.
 The analysis is limited to the standard continuous trading sessions, from 09:00 to 17:30 on Xetra and from 09:00 to 17:25 on Nasdaq Stockholm, expressed in local exchange time.

\begin{figure}[h]
    \begin{subfigure}{.5\textwidth}
        \includegraphics[width=0.92\linewidth]{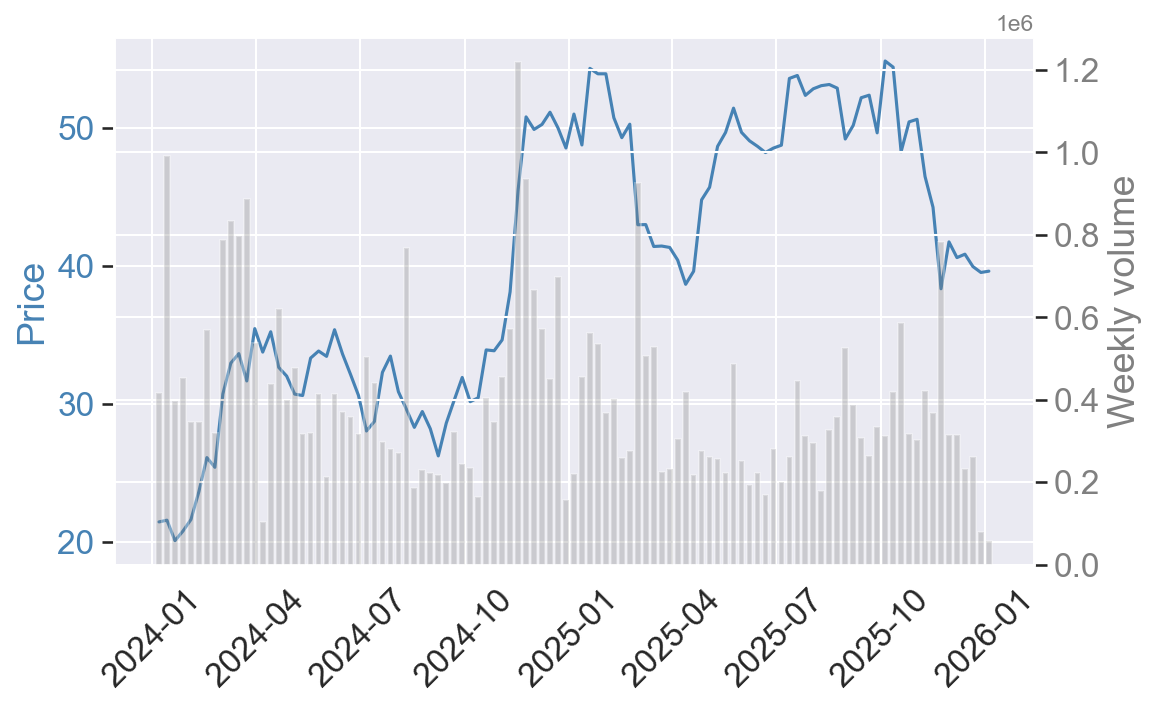}
        \caption{VBTC.XE}
    \end{subfigure}
    \begin{subfigure}{.5\textwidth}
        \includegraphics[width=0.92\linewidth]{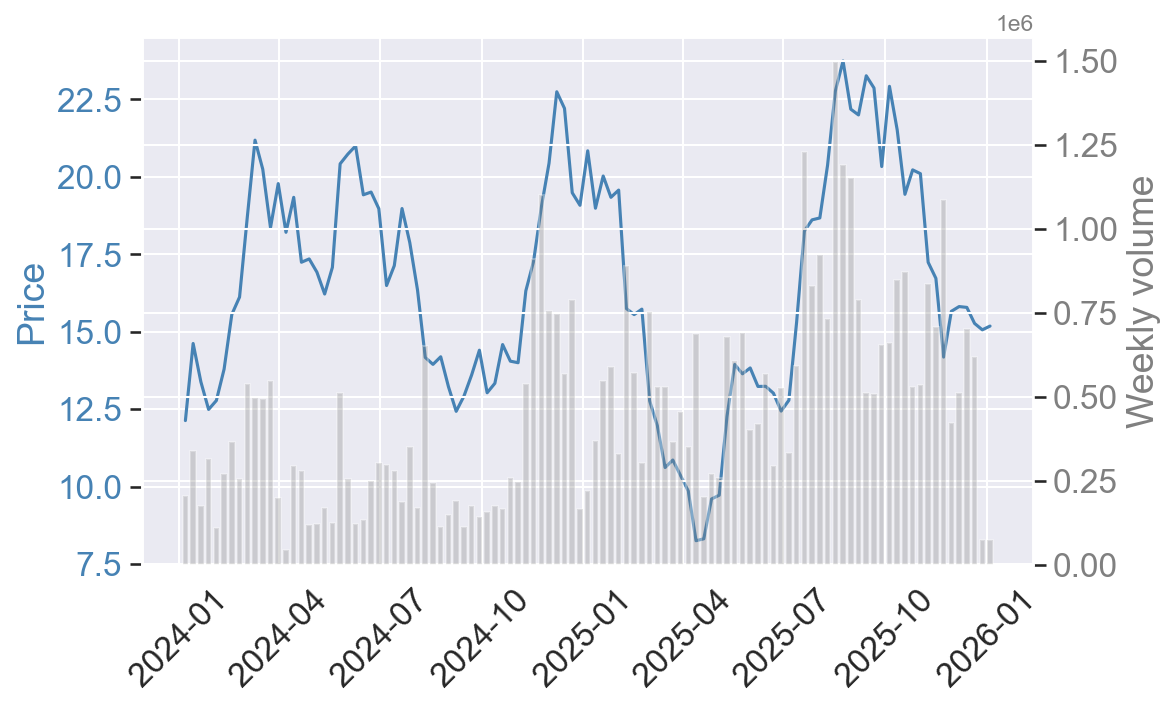}
        \caption{VETH.XE}
    \end{subfigure}
    \begin{subfigure}{.5\textwidth}
        \includegraphics[width=0.92\linewidth]{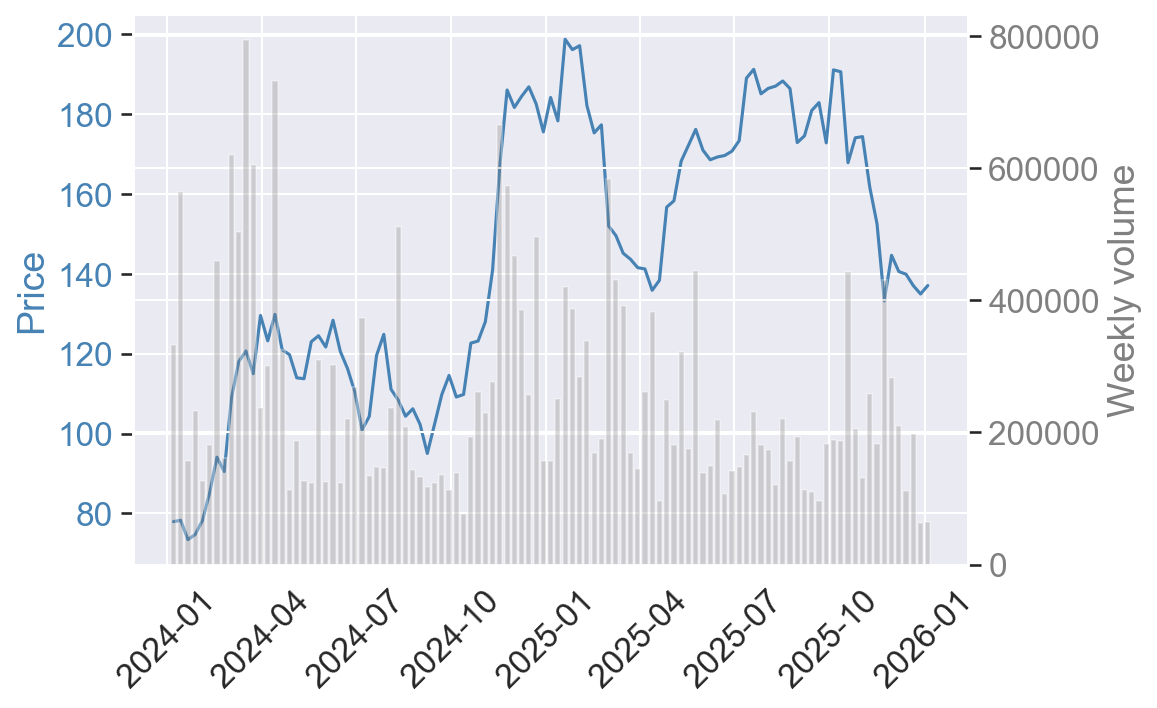}
        \caption{VIRBTC.ST}
    \end{subfigure}
    \begin{subfigure}{.5\textwidth}
        \includegraphics[width=0.92\linewidth]{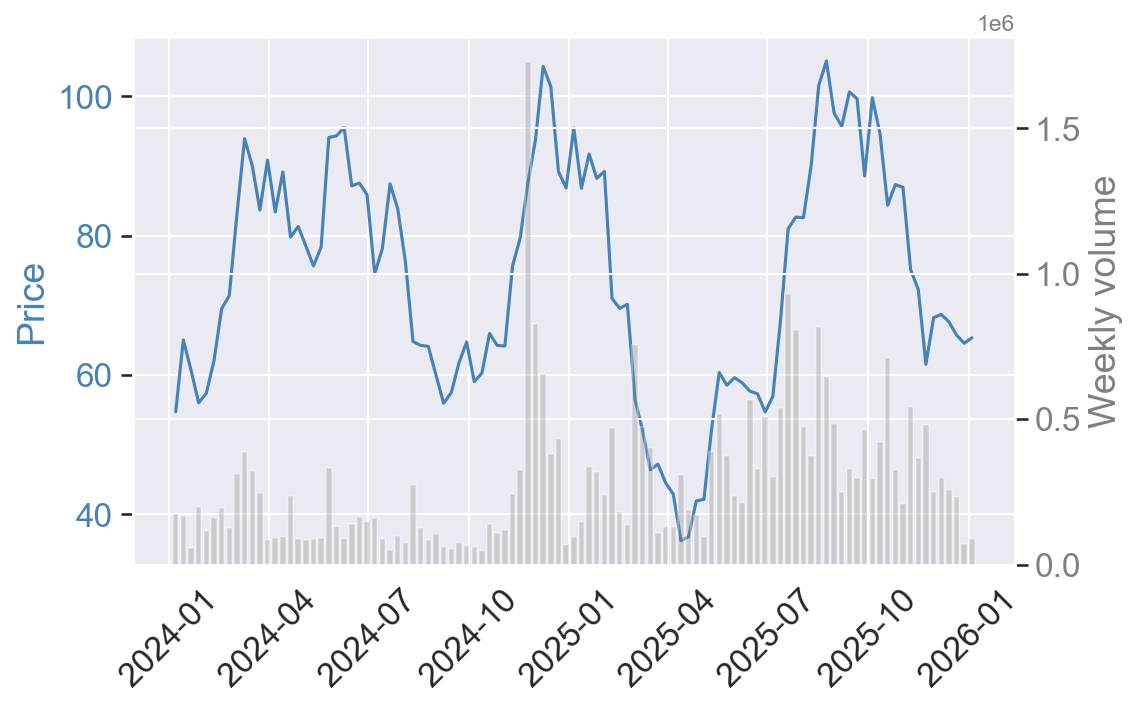}
        \caption{VIRSETHS.ST}
    \end{subfigure}
  \caption{Weekly closing price and weekly aggregate trade volume for the four cryptocurrency ETPs over the sample period. Price is defined as the last available one-minute closing price within each calendar week. Volume is the sum of all one-minute trade volumes within the week, expressed in units traded.}
    \label{fig:price_volume}
\end{figure}

In Figure~\ref{fig:price_volume} we present the weekly closing price and weekly 
aggregate trade volume for the four instruments over the sample period. All 
four products show a similar price trajectory. Prices fell during the first half of 2024, rose sharply in the fourth quarter, and reached their highest levels in early 2025 before declining again. These changes follow the price movements of Bitcoin and Ethereum over the same period.

For all four instruments, higher trading volume is associated with larger price movements.
 Trading volume is highest in late 2024 and early 2025, when prices also reach their highest levels.
 VBTC.XE has the highest trading volume among the four instruments, making it the most liquid ETP in the sample. On the other hand, VIRSETHS.ST has the lowest and most volatile trading volume over the sample period.

\begin{table}[h]
\centering
\caption{Summary statistics of minute-level log-returns (\%) for the four cryptocurrency ETPs over the sample period January 2024 -- December 2025.}
\label{tab:ret_stats}
\begin{tabular}{lrrrr}
\toprule
Statistic        & VBTC.XE  & VETH.XE  & VIRBTC.ST & VIRSETHS.ST \\
\midrule
Mean             & $-0.00029$ & $-0.00167$ & $-0.00094$ & $-0.00380$ \\
Std.\ dev.\      & 0.1229    & 0.1939    & 0.2361    & 0.3698      \\
Min              & $-4.585$  & $-5.024$  & $-5.391$  & $-14.133$   \\
$p_1$            & $-0.334$  & $-0.530$  & $-0.518$  & $-0.990$    \\
$p_5$            & $-0.183$  & $-0.280$  & $-0.295$  & $-0.510$    \\
$p_{95}$         & 0.181     & 0.278     & 0.299     & 0.495       \\
$p_{99}$         & 0.334     & 0.518     & 0.527     & 1.012       \\
Max              & 3.722     & 5.092     & 5.698     & 6.556       \\
Skewness         & $-0.196$  & 0.024     & 0.278     & $-0.652$    \\
Excess kurtosis  & 39.92     & 27.55     & 57.34     & 58.48       \\
\bottomrule
\end{tabular}
\end{table}

We begin by examining the basic statistical properties of the analysed series, summarised in Table~\ref{tab:ret_stats}.
Average one-minute log-returns are close to zero for all four instruments, ranging from $-0.00029\%$ for VBTC.XE to $-0.00380\%$ for VIRSETHS.ST.
Volatility increases monotonically from 
VBTC.XE (0.123\%) through VETH.XE (0.194\%) and VIRBTC.ST (0.236\%) 
to VIRSETHS.ST (0.370\%), reflecting both the higher underlying volatility 
of Ethereum relative to Bitcoin and the lower liquidity of the 
Stockholm-listed instruments.
For the excess kurtosis we observe values ranging from 27.55 for VETH.XE to 58.48 for VIRSETHS.ST.

We fit five candidate distributions to the minute-level log-returns of each 
instrument: the normal inverse Gaussian (NIG), the generalised normal, 
Student's $t$, the non-central $t$ (NCT), and the double Weibull \citep{barndorff1977exponentially, burnecki2011stability, hogben1961moments, konczal2026pricing, nadarajah2005generalized}. 
We evaluate goodness-of-fit using the Kolmogorov-Smirnov (KS) and the 
Anderson-Darling (AD) tests \citep{smirnov1948table, anderson1954test}. The AD statistic is given particular weight 
in our assessment, as it places greater emphasis on tail fit relative to the 
KS statistic, which is more sensitive to discrepancies in the centre of the 
distribution.

Both the KS and AD tests strongly rejected all candidate distributions, with p-values below 0.001 in every case.
 This result may be related to the very large sample size, because goodness-of-fit tests become highly sensitive even to small deviations from the assumed distribution. Nevertheless, based on the AD test statistic, the NIG distribution provided the best fit for three instruments, while the double Weibull provided the best fit for VIRSETHS.ST.

\begin{figure}[h]
		\begin{subfigure}{.5\textwidth}
			\centering
			\includegraphics[width=1\linewidth]{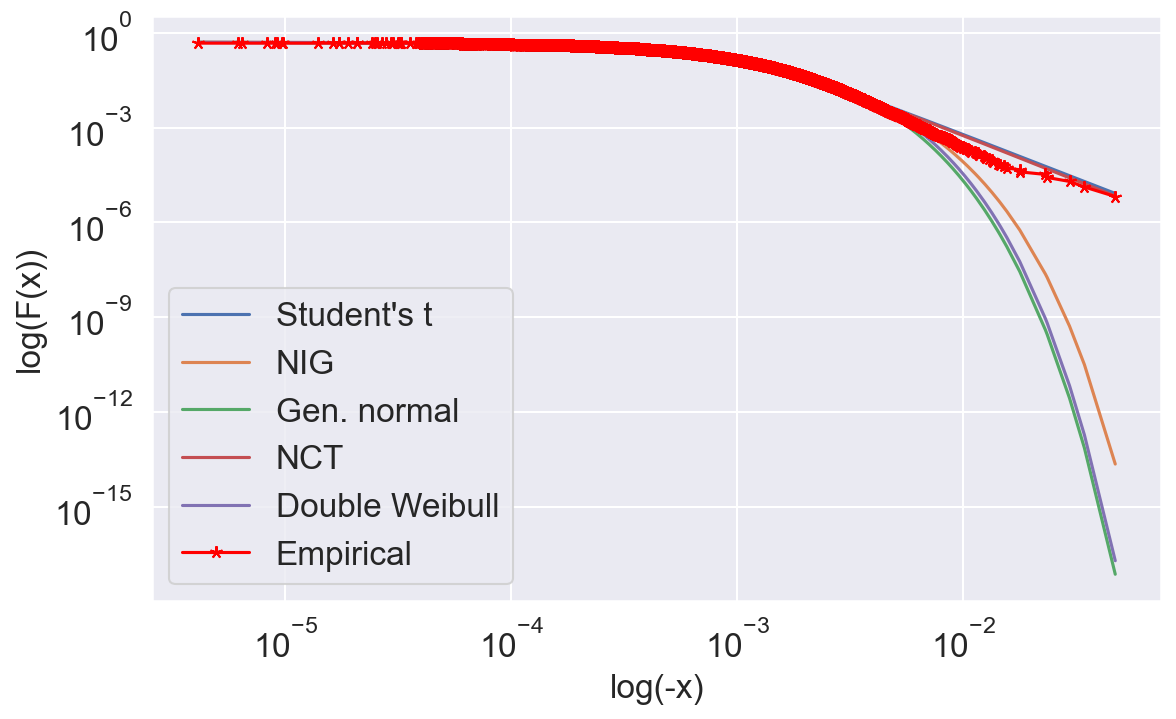}  
			\caption{VBTC.XE}
			\label{fig:lewyogonVBTCXE}
		\end{subfigure}
		\hspace{1em}
		\begin{subfigure}{.5\textwidth}
			\centering
			\includegraphics[width=1\linewidth]{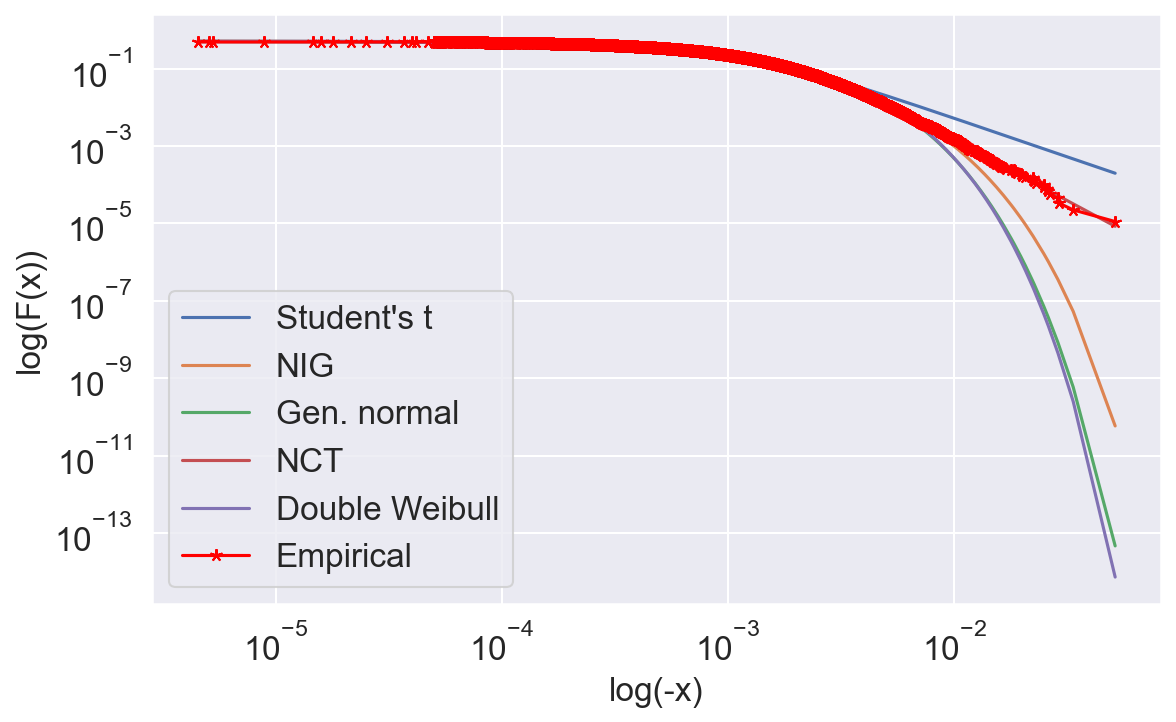}  
			\caption{VETH.XE}
			\label{fig:lewyogonVETHXE}
		\end{subfigure}
		\begin{subfigure}{.5\textwidth}
			\centering
			\includegraphics[width=1\linewidth]{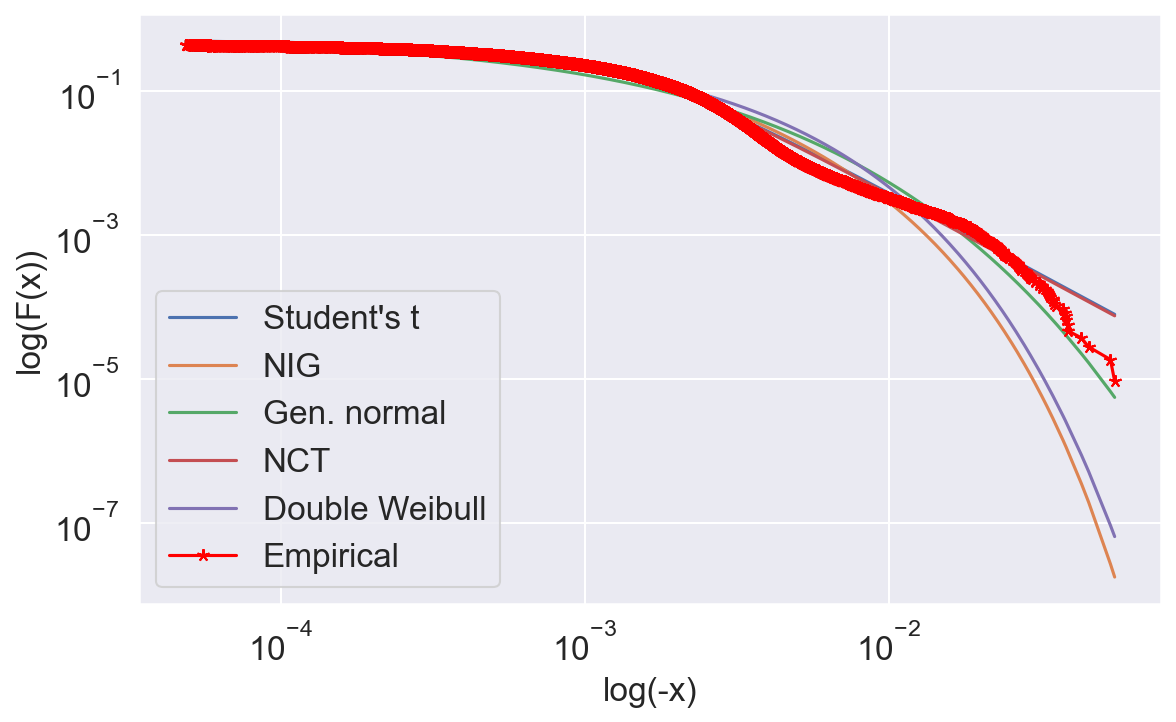}  
			\caption{VIRBTC.ST}
			\label{fig:lewyogonVIRBTCST}
		\end{subfigure}
		\hspace{1em}
		\begin{subfigure}{.5\textwidth}
			\centering
			\includegraphics[width=1\linewidth]{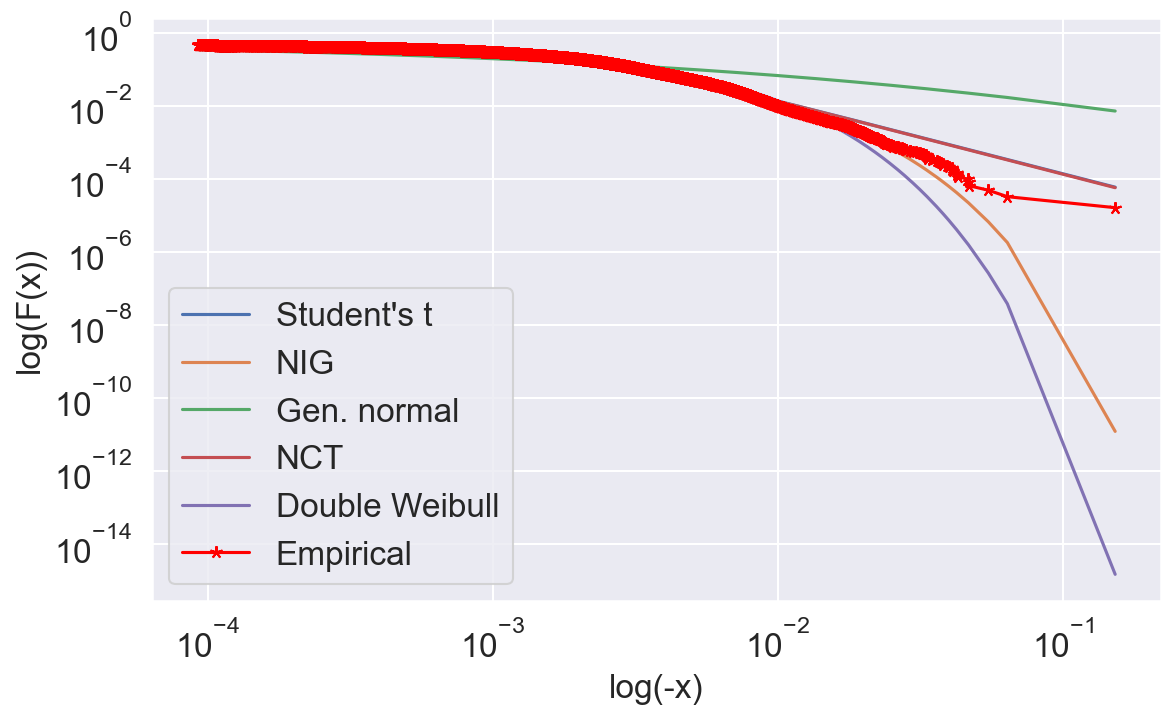}  
			\caption{VIRSETHS.ST}
			\label{fig:lewyogonVIRSETHSST}
		\end{subfigure}
		
		\caption{Left tails of the empirical and fitted analytical distribution functions. The plots are presented on the log-log scale.}
		\label{fig:lewyogon}
	\end{figure}

In Figure~\ref{fig:lewyogon} we illustrate the left tails of the empirical and fitted 
distribution functions on a log-log scale for all four instruments. 
Among the fitted distributions, the NCT and Student's $t$ provide the closest approximation to the 
empirical tail across most of the range. The generalised normal and 
double Weibull distributions underestimate tail probabilities.
We observe a similar pattern for all four instruments.

    \begin{figure}[h]
		\begin{subfigure}{.5\textwidth}
			\centering
			\includegraphics[width=1\linewidth]{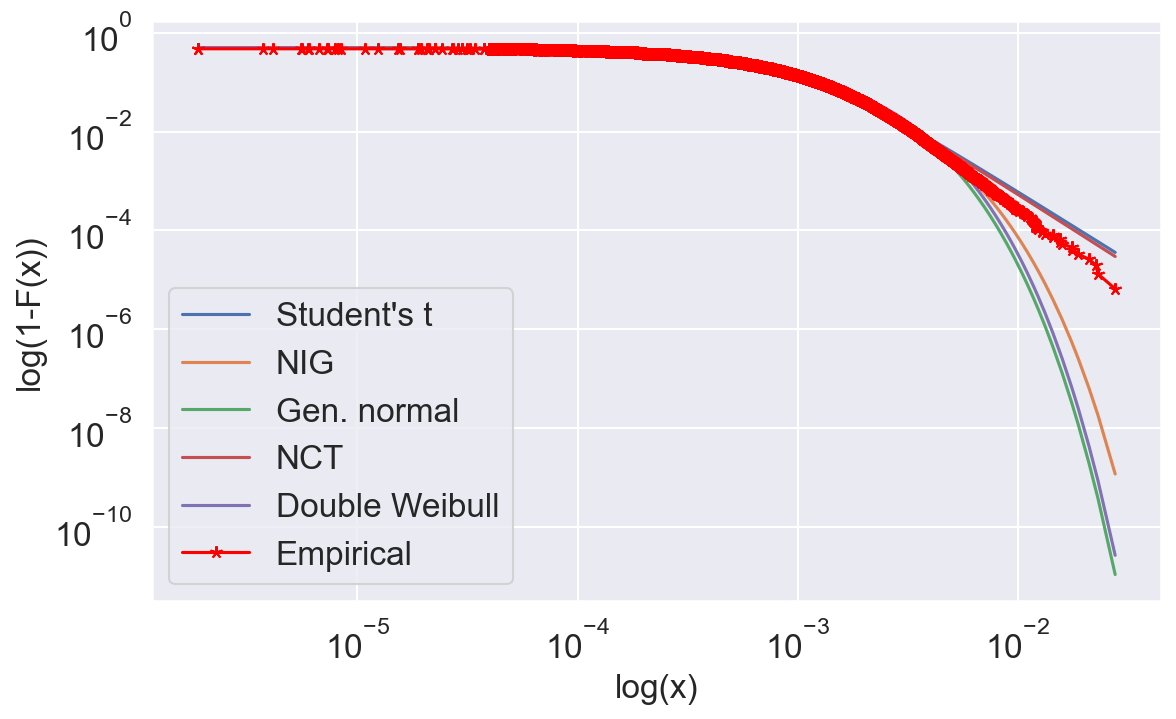}  
			\caption{VBTC.XE}
			\label{fig:prawyogonVBTCXE}
		\end{subfigure}
		\hspace{1em}
		\begin{subfigure}{.5\textwidth}
			\centering
			\includegraphics[width=1\linewidth]{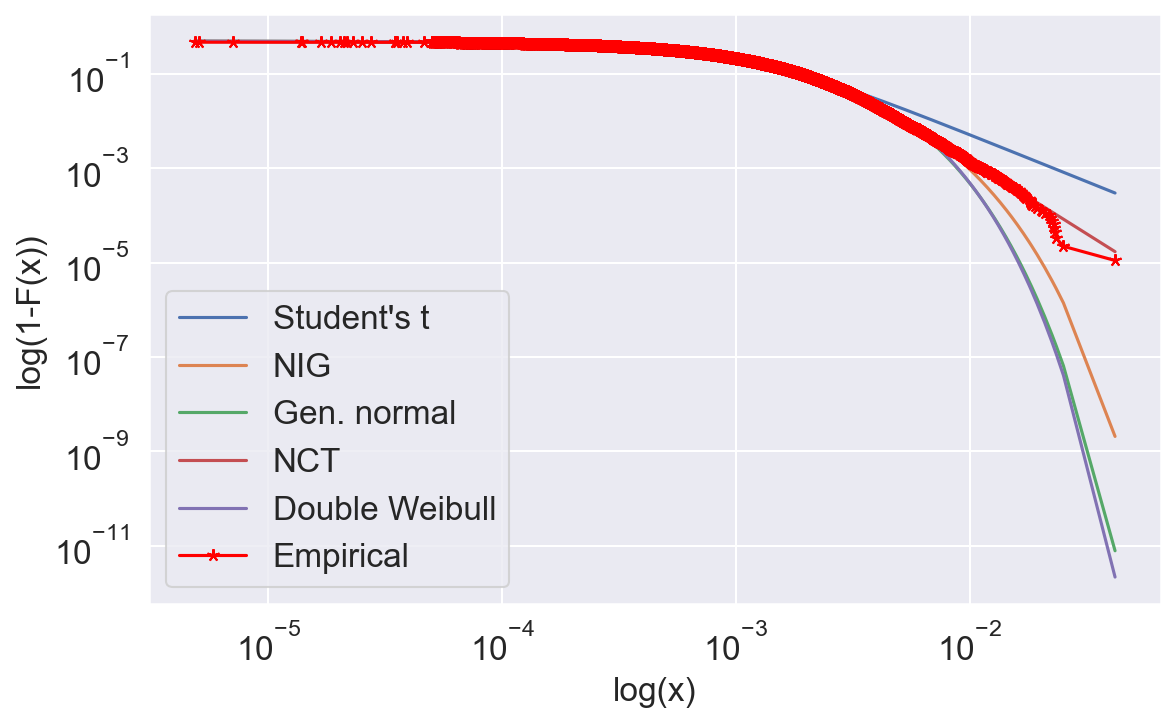}  
			\caption{VETH.XE}
			\label{fig:prawyogonVETHXE}
		\end{subfigure}
		\begin{subfigure}{.5\textwidth}
			\centering
			\includegraphics[width=1\linewidth]{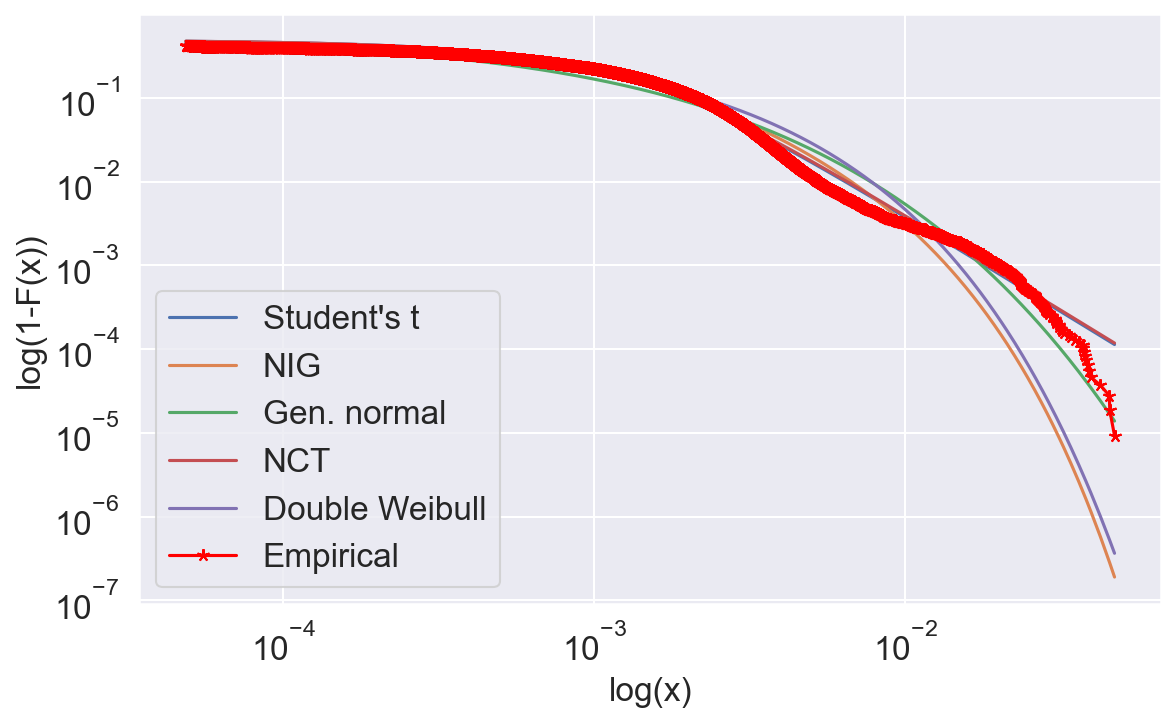}  
			\caption{VIRBTC.ST}
			\label{fig:prawyogonVIRBTCST}
		\end{subfigure}
		\hspace{1em}
		\begin{subfigure}{.5\textwidth}
			\centering
			\includegraphics[width=1\linewidth]{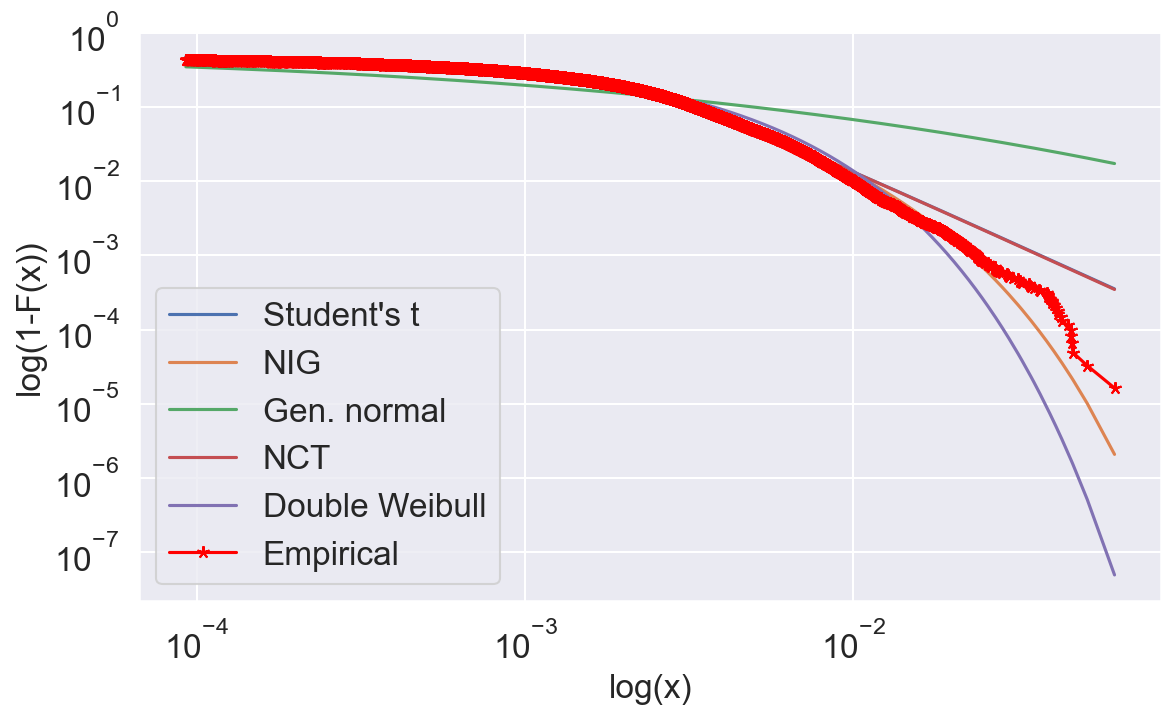}  
			\caption{VIRSETHS.ST}
			\label{fig:prawyogonVIRSETHSST}
		\end{subfigure}
		
		\caption{Right tails of the empirical and fitted analytical distribution functions. The plots are presented on the log-log scale.}
		\label{fig:prawyogon}
	\end{figure}

    The right tails, presented in Figure~\ref{fig:prawyogon}, show a similar pattern. Student's $t$ 
and NCT again provide the closest approximation across most of the tail 
range, while the generalised normal shows the earliest decline in performance. Notably, the 
right tails are better approximated by the parametric distributions 
than the left tails.

Overall, based on the AD statistic, NIG provides the best fit to the full return distribution for three instruments, while the double Weibull performs best for VIRSETHS.ST. However, the tail plots show that the NCT and Student's $t$ distributions follow the empirical tails more closely. This suggests that the preferred distribution depends on whether the full distribution or the tails are considered.

\begin{figure}[h]
    \begin{subfigure}{.5\textwidth}
        \includegraphics[width=\linewidth]{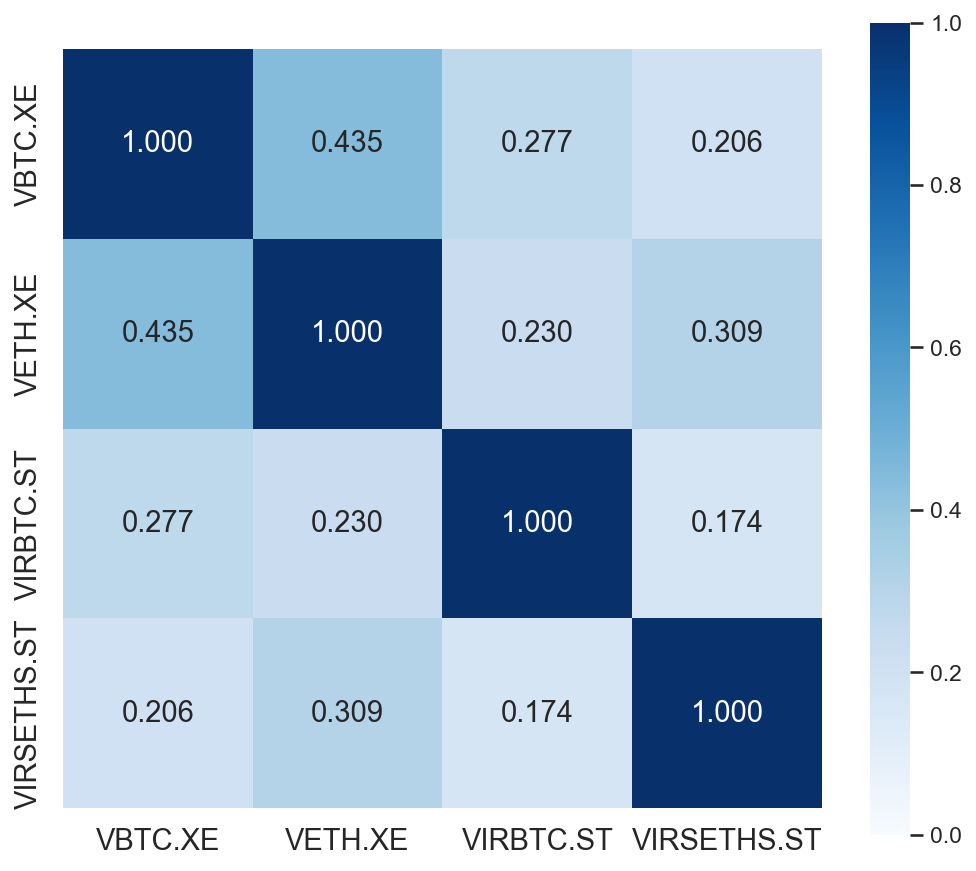}
        \caption{one-minute}
    \end{subfigure}
    \begin{subfigure}{.5\textwidth}
        \includegraphics[width=\linewidth]{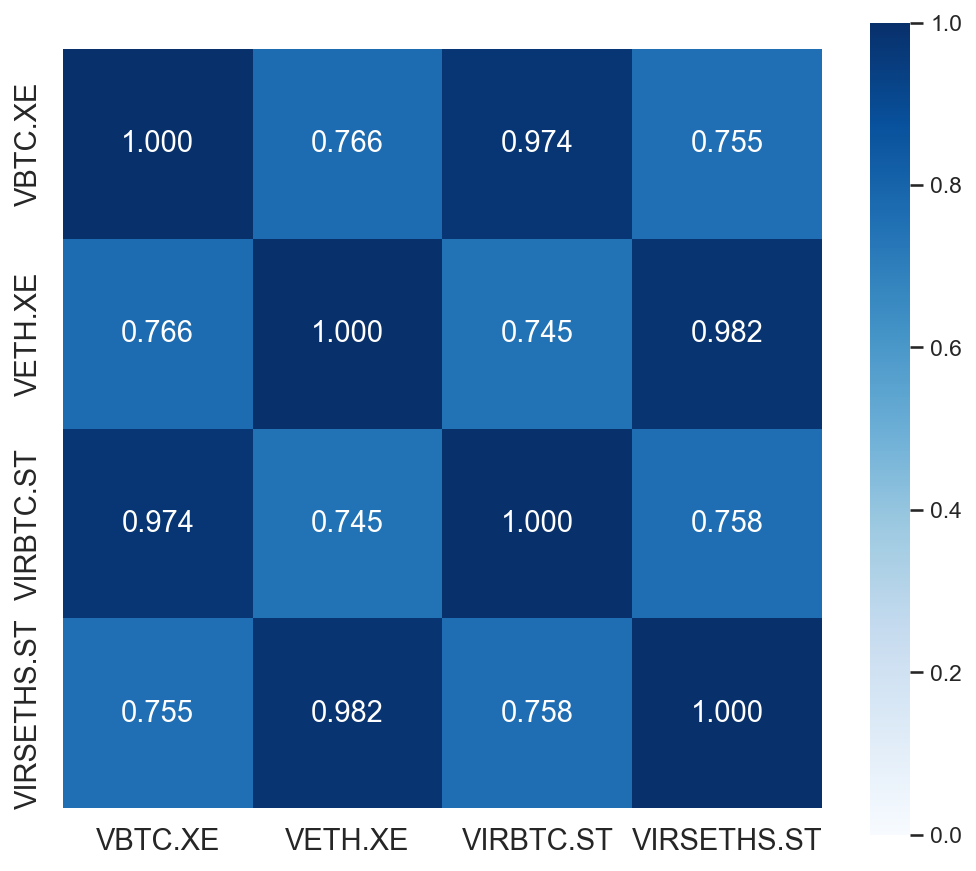}
        \caption{daily}
    \end{subfigure}
    \caption{Cross-symbol return correlations at the one-minute and daily frequency. }
    \label{fig:crosscorrelation}
\end{figure}

We compare correlations between the four selected instruments to check how their relationships change with the sampling frequency. Figure~\ref{fig:crosscorrelation} presents Pearson correlations calculated using one-minute and daily returns. We observe differences in the calculated values between the two sampling frequencies. At the one-minute frequency, correlations between the instruments are relatively low, ranging from 0.174 to 0.435. The strongest correlations are found between instruments traded on the same venue.

This lower correlation at shorter sampling intervals is known as the Epps effect and is widely documented in market microstructure research 
\citep{epps1979comovements}. The Epps effect describes the decrease in return correlations observed at shorter sampling intervals. This effect is commonly explained by asynchronous trading, bid–ask bounce, and discrete price changes in high-frequency data.
At the daily frequency correlations increase across all pairs, 
with values ranging from 0.745 to 0.982.

\begin{figure}[h]
    \begin{subfigure}{.5\textwidth}
        \includegraphics[width=\linewidth]{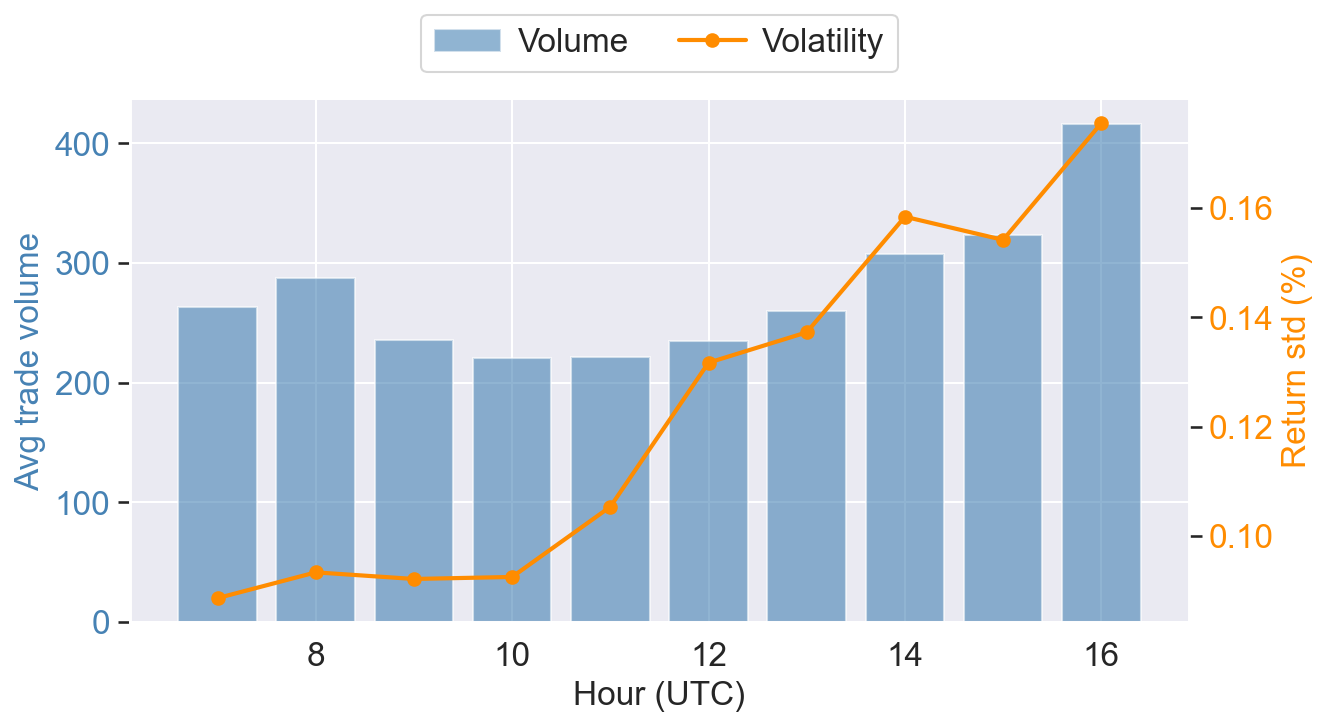}
        \caption{VBTC.XE}
    \end{subfigure}
    \begin{subfigure}{.5\textwidth}
        \includegraphics[width=\linewidth]{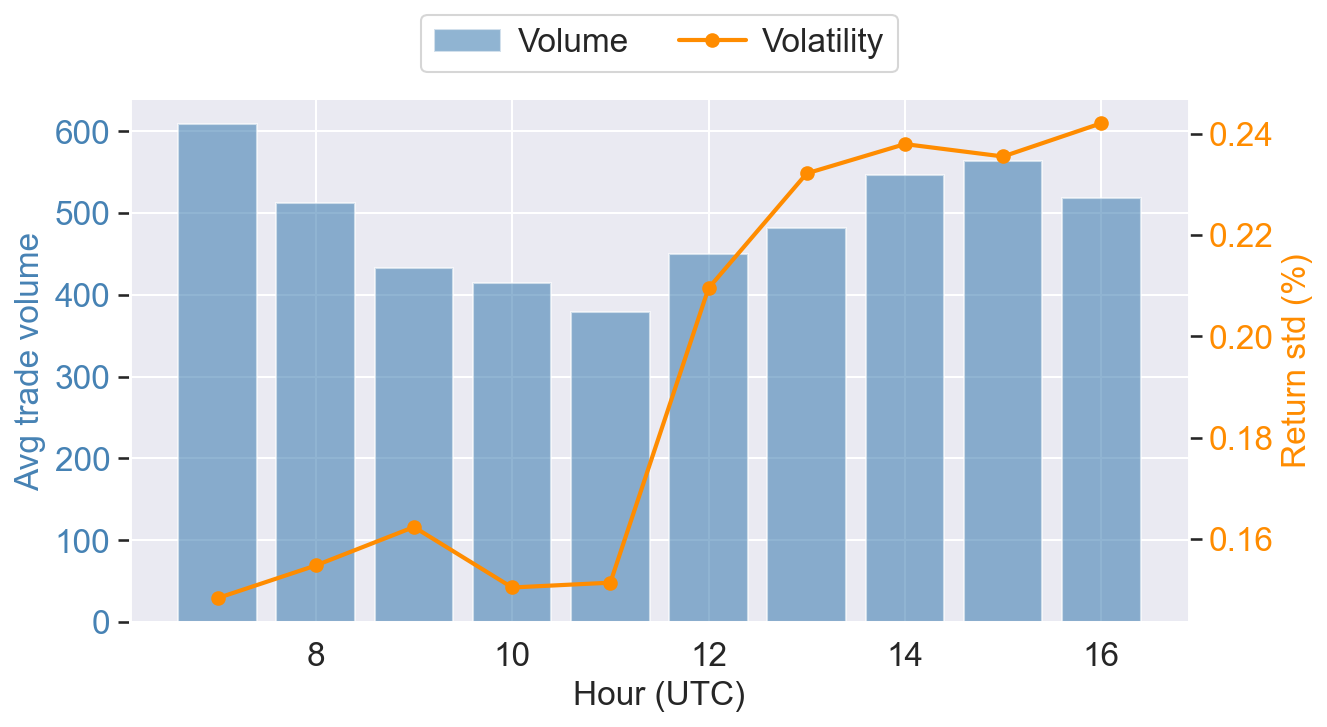}
        \caption{VETH.XE}
    \end{subfigure}
    \begin{subfigure}{.5\textwidth}
        \includegraphics[width=\linewidth]{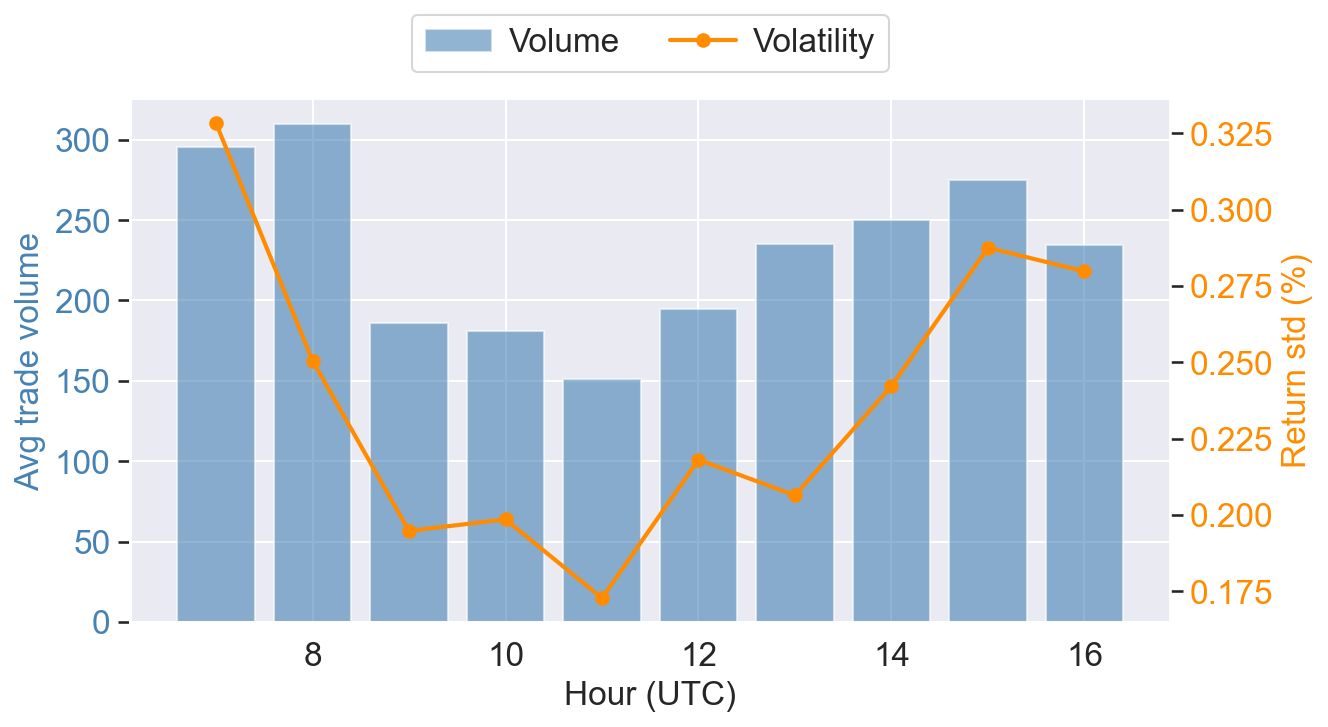}
        \caption{VIRBTC.ST}
    \end{subfigure}
    \begin{subfigure}{.5\textwidth}
        \includegraphics[width=\linewidth]{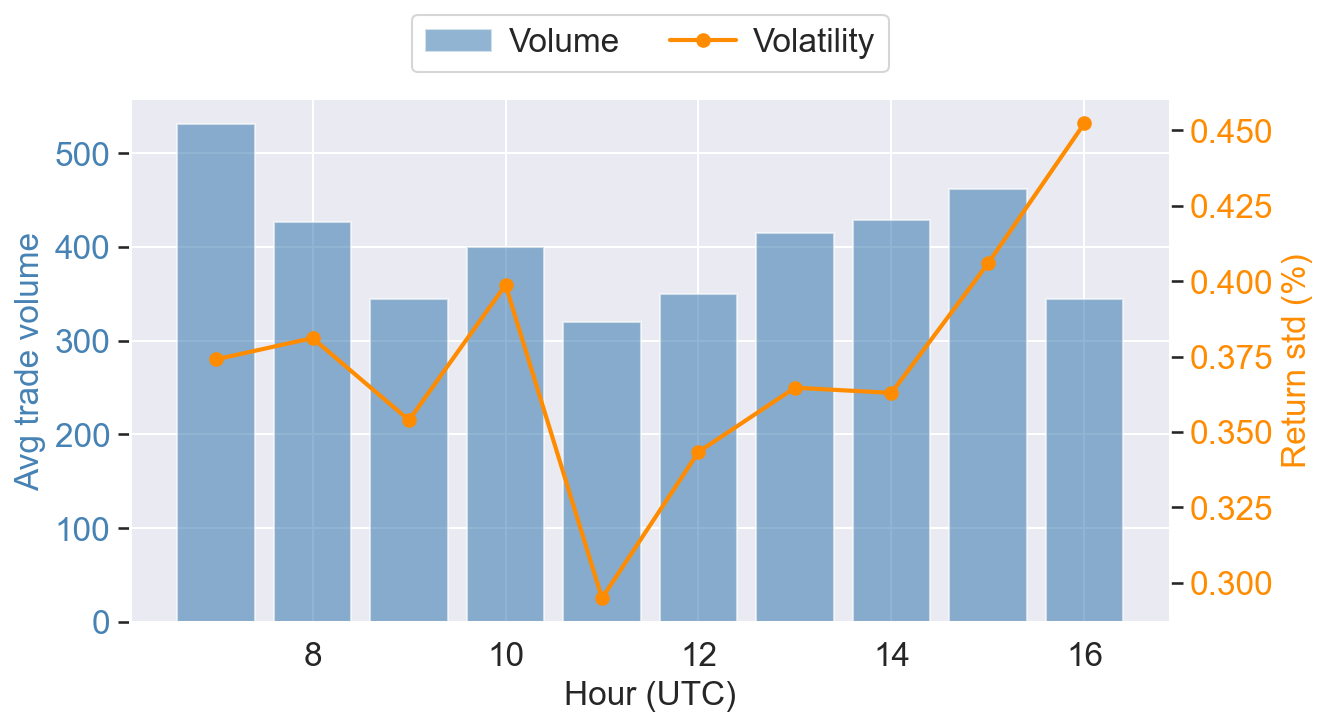}
        \caption{VIRSETHS.ST}
    \end{subfigure}
    \caption{Intraday volume and volatility patterns during exchange hours, averaged over the full sample period. Bars show hourly mean trade volume (left axis); the connected markers show the hourly standard deviation of one-minute returns (right axis, in \%).}
    \label{fig:intraday_patterns}
\end{figure}

Having examined how the relationships between instruments vary with sampling frequency, we now turn to the intraday patterns within each individual instrument.
In Figure~\ref{fig:intraday_patterns} we present the intraday pattern of trading volume and return volatility for the four instruments, both measured hourly and averaged over the full sample period. For the two Xetra instruments, volatility increases throughout the trading day, from approximately 0.09\% at the open to 0.175\% in the final hour for VBTC.XE, and from around 0.15\% to 0.24\% for VETH.XE, with no opening spike on either instrument. This pattern differs from the classical U-shaped intraday volatility profile documented in equity markets \citep{harris1986transaction, wood1985investigation}. However, trading volume follows a different pattern for the two instruments. For VBTC.XE, average trading volume remains stable during most of the session but rises sharply in the final hour, together with volatility. This end-of-day increase may reflect the closing auction effect and the concentration of institutional trading near the market close \citep{admati1988theory, kyle1985continuous}. VETH.XE instead shows the highest volume at the open, a decline through the mid-session, and a partial recovery towards the close -- a U-shape that is more typical of instruments where information-driven trading dominates at the start of the day, and one that only partly tracks the instrument's monotonically rising volatility.

The instruments traded in Stockholm show a clearly different pattern.
 Both VIRBTC.ST and VIRSETHS.ST exhibit a pronounced volatility spike at the open, with standard deviations of approximately 0.31\% and 0.38\% respectively, followed by a sharp decline through the mid-morning; volatility then remains relatively stable through the middle of the session before rising again towards the close. Trading volume shows a similar pattern for both instruments, with high activity at the market open, a clear decline around midday, and a partial increase in the afternoon.
 For VIRBTC.ST, the minimum average volume of approximately 150 units occurs around noon, rising again to around 270 units by the close, while VIRSETHS.ST follows a comparable trajectory at higher absolute volume levels throughout the session. The joint elevation of volume and volatility at the open is consistent with the resolution of overnight information at the start of the trading session, and their common decline suggests that price discovery on Nasdaq Stockholm is largely completed within the first hour of trading. Kruskal--Wallis tests \citep{kruskal1952use} reject the null hypothesis of
equal distributions across hours of the trading day for both volume and
volatility, for all four instruments. 

\section{Anomaly detection}
\label{sec:anomaly}

After examining the return distributions and intraday volatility patterns of the four ETPs, we now focus on identifying unusual price movements.
We define a market anomaly as a one-minute bar in which the log-return
of a cryptocurrency ETP satisfies one or more formally specified
conditions indicative of abnormal behaviour. All definitions
produce a binary indicator $A_t \in \{0, 1\}$, where $A_t = 1$ denotes
an anomalous bar. We consider four definitions: a baseline definition
grounded in EVT ($D^{POT}$) and three novel definitions
capturing cross-venue divergence ($D^{CV}$), no recovery
($D^{NR}$), and momentum reversal ($D^{MR}$).

The baseline anomaly definition is grounded in the POT method \citep{davison1990models, 
mcneil2000estimation, embrechts1997modelling}. Let $r_t$ denote the one-minute
log-return at bar $t$ and let $L_t = -r_t$ denote the corresponding loss,
so that the left tail of the return distribution corresponds to the right
tail of the loss distribution. For a threshold $u$ below the right endpoint
$x_F$ of the loss distribution $F$, the excess distribution function is
$F_u(x) = P(L - u \leq x \mid L > u)$. By the Pickands--Balkema--de Haan
theorem \citep{balkema1974residual, pickands1975statistical}, $F$ belongs
to the maximum domain of attraction of an extreme value distribution if
and only if
\begin{equation}
    \lim_{u \to x_F} \; \sup_{0 \le x < x_F - u}
\bigl| F_u(x) - G_{\xi,\beta(u)}(x) \bigr| = 0
\end{equation}
for some positive measurable function $\beta(u)$, where
$G_{\xi,\beta}$ is the GPD
\begin{equation}
G_{\xi,\beta}(x) =
\begin{cases}
1 - \left(1 + \dfrac{\xi x}{\beta}\right)^{-1/\xi} & \xi \neq 0, \\[6pt]
1 - e^{-x/\beta} & \xi = 0,
\end{cases}
\end{equation}
with scale $\beta > 0$ and $x \geq 0$ for $\xi \geq 0$ as well as
$0 \leq x \leq -\beta/\xi$ for $\xi < 0$.

We set the threshold $\hat{u}$ as the 95th percentile of the training-sample
losses, equivalently $-\hat{u} = \hat{q}_{0.05}(r)$ is the 5th percentile of
training returns, so that the empirical exceedance rate is
$\hat{\zeta}_u = n_u/N_{\mathrm{train}} \approx 0.05$. A GPD is fitted by maximum likelihood to the exceedances $L_t-\hat{u}$, giving estimates $\hat{\xi}$ and $\hat{\beta}$ of the shape and scale parameters, respectively.
The anomaly threshold corresponding to a target unconditional anomaly rate $p=0.01$ is obtained by inverting the tail estimator
$\widehat{P}(L>q)=\hat{\zeta}_u
\bigl(1-G_{\hat{\xi},\hat{\beta}}(q-\hat{u})\bigr)$,
which gives
\begin{equation}
\hat{q}_p =
\begin{cases}
\hat{u} + \dfrac{\hat{\beta}}{\hat{\xi}}
\left[
\left(\dfrac{p}{\hat{\zeta}_u}\right)^{-\hat{\xi}} - 1
\right],
& \hat{\xi} \neq 0, \\
\hat{u} + \hat{\beta}
\log\left(\dfrac{\hat{\zeta}_u}{p}\right),
& \hat{\xi} = 0.
\end{cases}
\end{equation}
Bar $t$ is classified as a $D^{POT}$ anomaly if
\begin{equation}
D^{POT}_t = \mathbf{1}\!\left[\, r_t < -\hat{q}_p \,\right].
\end{equation}
Since $\hat{q}_p$ is estimated from the training sample but applied to the entire dataset, the observed full-sample anomaly rate may not be exactly equal to the target probability $p$.

We use the POT-based definition as a benchmark, since it identifies unusually large negative returns within each instrument separately.
The economic phenomena captured by the three additional indicators have already been discussed in the financial literature.
 Our contribution is to express these phenomena as explicit binary anomaly indicators designed for high-frequency cryptocurrency ETP data.

The POT approach does not use the cross-venue dimension of our dataset. We therefore introduce a second definition based on the idea that products linked to the same underlying cryptocurrency should display similar return behaviour across venues. When the returns of two products tracking the same cryptocurrency temporarily differ, this may indicate a short-term pricing difference between the exchanges rather than a movement in the underlying cryptocurrency.
 Related cross-market price discrepancies have been studied for homogeneous or closely linked securities traded in multiple markets and in the pairs-trading literature \citep{eun2003cross, elliott2005pairs}. We use this approach to define a binary anomaly indicator based on rolling residuals for ETPs that track the same cryptocurrency on different exchanges.
 Let $r^{(1)}_t$ and $r^{(2)}_t$ denote the one-minute log-returns of two ETPs tracking the same cryptocurrency on Xetra and Nasdaq Stockholm, respectively. We use only bars in which both instruments are actively traded.
Because the two listings need not
respond one-for-one to movements in the underlying asset, we estimate
the systematic sensitivity between them by a rolling ordinary least
squares regression without intercept
\begin{equation}
    r^{(2)}_s = \beta_t\, r^{(1)}_s + \varepsilon_s,
\qquad s = t-n, \ldots, t-1,
\end{equation}
over a window of $n = 1440$ preceding active bars, corresponding to
approximately three full trading sessions. The residual $\hat{\varepsilon}_t = r^{(2)}_t - \hat{\beta}_t r^{(1)}_t$ measures the part of the return of the second listing that cannot be explained by the return of the first listing.
 Bar $t$
is classified as a $D^{CV}$ anomaly if
\begin{equation}
   D^{CV}_t = \mathbf{1}\!\left[
    |\hat{\varepsilon}_t| >
    \hat{q}_{1-p}\!\left(
        \bigl\{|\hat{\varepsilon}_s|\bigr\}_{s=t-n}^{t-1}
    \right)
\right], 
\end{equation}
where $p = 0.01$. We calculate the residual in one fixed direction and use the same anomaly label for both instruments in each pair. Therefore, the same bars are marked as anomalies for both listings.

We next introduce a third definition that considers both the size of the initial price movement and what happens immediately afterwards.
 The motivation is that large negative returns may represent different types of events: some sharp drops reverse quickly, while others are followed by little or no recovery.
 Previous studies have analysed how quickly and to what extent prices recover after mini flash crashes and extreme negative intraday returns \citep{braun2018impact}.
We express this concept as a binary no-recovery indicator defined by a fixed forward horizon and a recovery threshold.
The no-recovery anomaly identifies extreme declines followed by limited recovery by the end of the selected observation window.
A flash crash is a sharp price drop followed by a quick recovery \citep{kirilenko2017flash}, while a lasting correction is followed by little or no recovery.
 Let
$S^K_t = \sum_{j=1}^{K} r_{t+j}$ denote the cumulative return over
the $K$ bars following bar $t$. Bar $t$ is classified as a $D^{NR}$
anomaly if
\begin{equation}
   D^{NR}_t = \mathbf{1}\!\left[
    r_t < \hat{q}_{p}(r,\,[t-n,\,t-1])
    \;\wedge\;
    S^K_t < c\,|r_t|
\right], 
\end{equation}
where $n = 1440$, $p = 0.01$, and $\hat{q}_{p}(r,\,[t-n,\,t-1])$ is the
empirical percentile of returns over the $n$ preceding active bars.
The forward window has length $K=10$, meaning ten active bars. When trading is continuous, this corresponds to ten minutes. The recovery threshold is set to $c=0.3$.
The condition $S_t^K<c|r_t|$ means that, by the end of the $K$-bar observation window, the cumulative log-return following the initial decline is less than $30\%$ of the absolute initial log-return. This definition measures recovery at the end of the window.
Unlike the other definitions,
$D^{NR}_t$ depends on returns realised after bar $t$ and is therefore
only observable with a delay of $K$ bars.

Momentum and return reversals have been widely documented in financial markets \citep{jegadeesh1993returns, lehmann1990fads}. Based on this literature, we define a binary momentum-reversal indicator for extreme negative returns that follow positive short-term momentum.
 Let $M_t = \sum_{j=1}^{5} r_{t-j}$ denote the short-term
momentum over the five bars preceding bar $t$. Bar $t$ is classified
as a $D^{MR}$ anomaly if
\begin{equation}
   D^{MR}_t = \mathbf{1}\!\left[
    r_t < \hat{q}_{p}(r,\,[t-n,\,t-1])
    \;\wedge\;
    M_t > 0
\right], 
\end{equation}
where $n = 1440$ and $p=0.01$. The condition $M_t>0$ means that the cumulative return over the five bars immediately before bar $t$ is positive. Thus, the extreme negative return at bar $t$ reverses the preceding upward price movement.

\begin{figure}[h]
    \begin{subfigure}{.45\textwidth}
        \includegraphics[width=\linewidth]{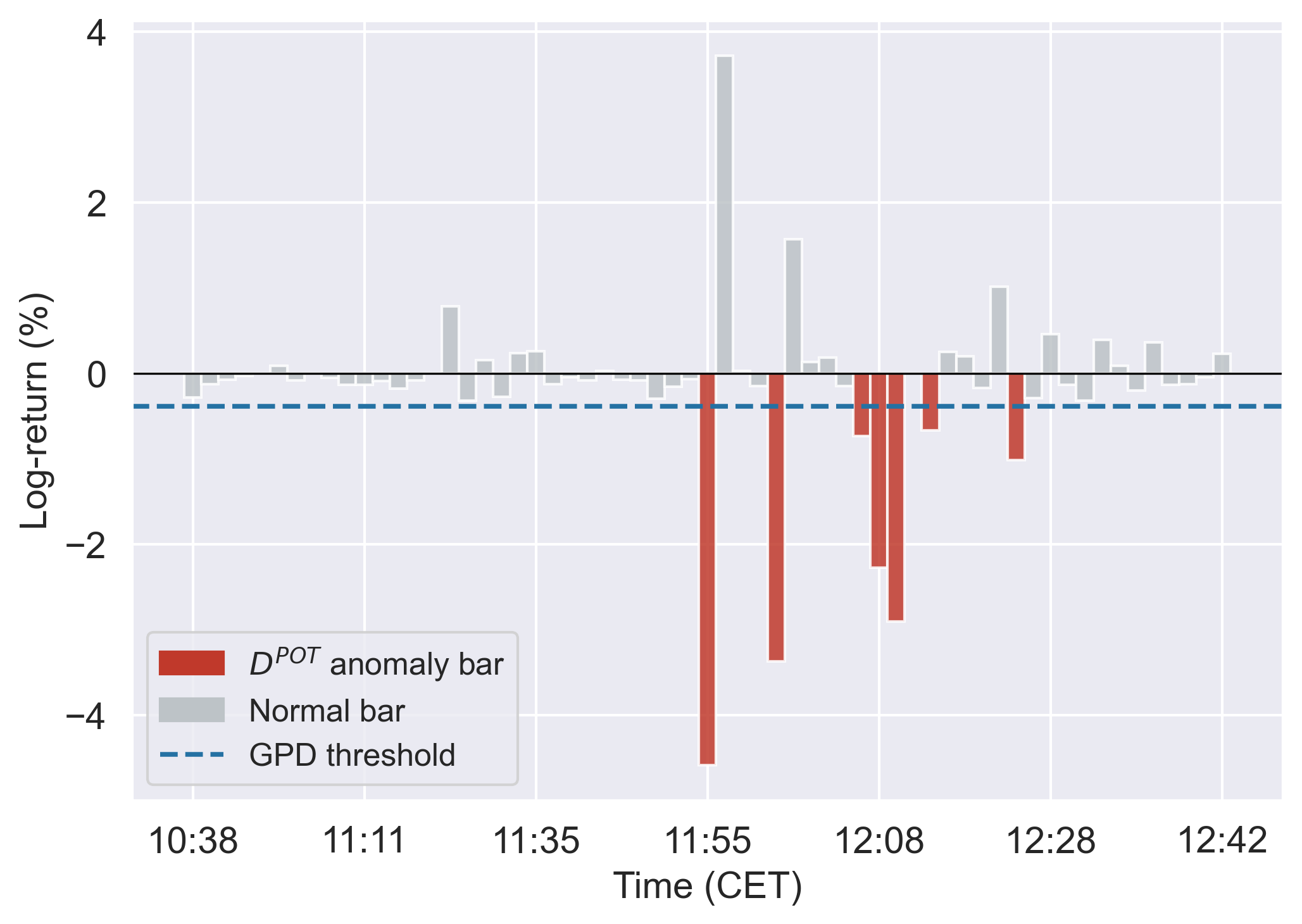}
        \caption{$D^{POT}$}
    \end{subfigure}
    \hfill
    \begin{subfigure}{.45\textwidth}
        \includegraphics[width=\linewidth]{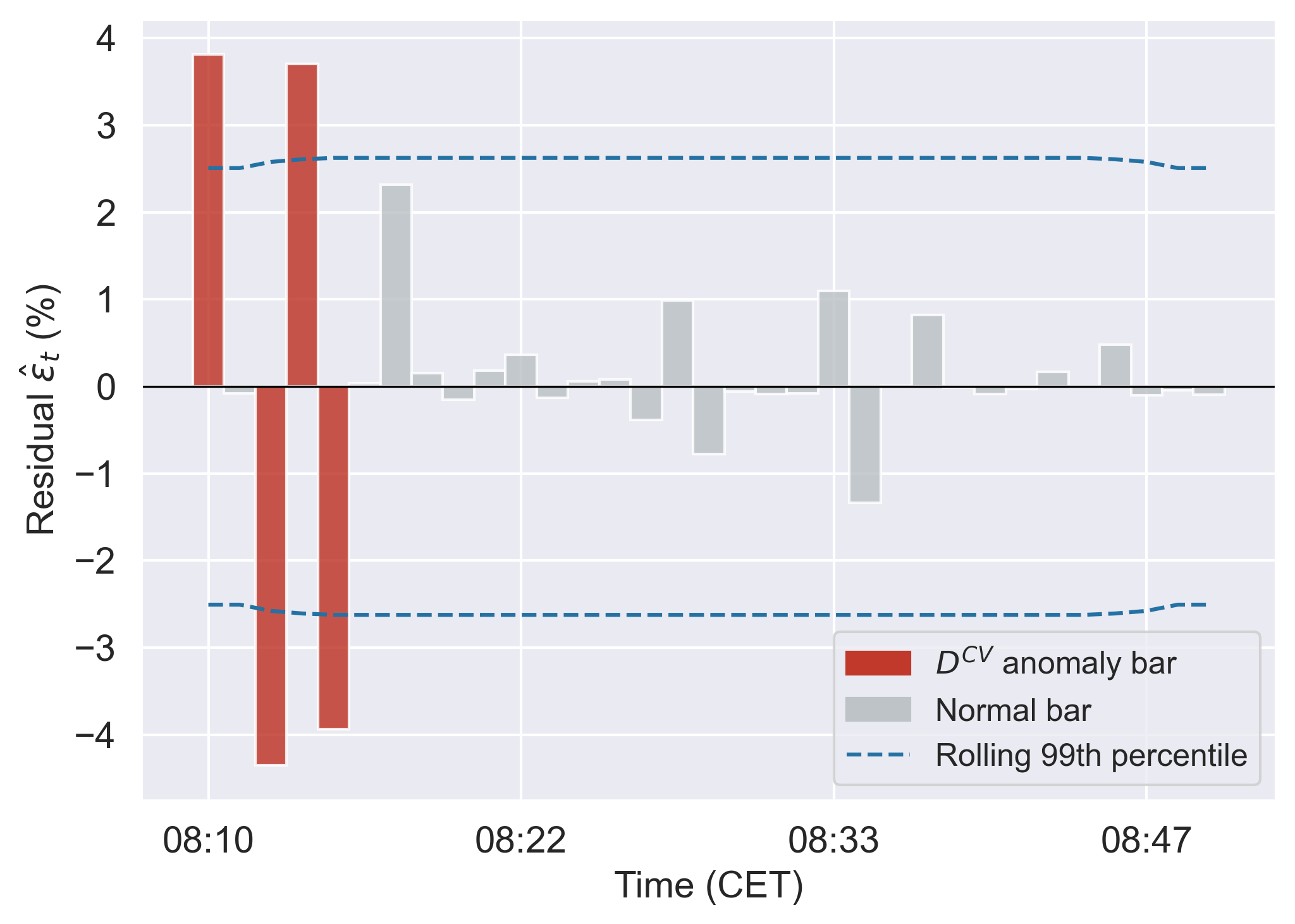}
        \caption{$D^{CV}$}
    \end{subfigure}
    \\[1em]
    \begin{subfigure}{.45\textwidth}
        \includegraphics[width=\linewidth]{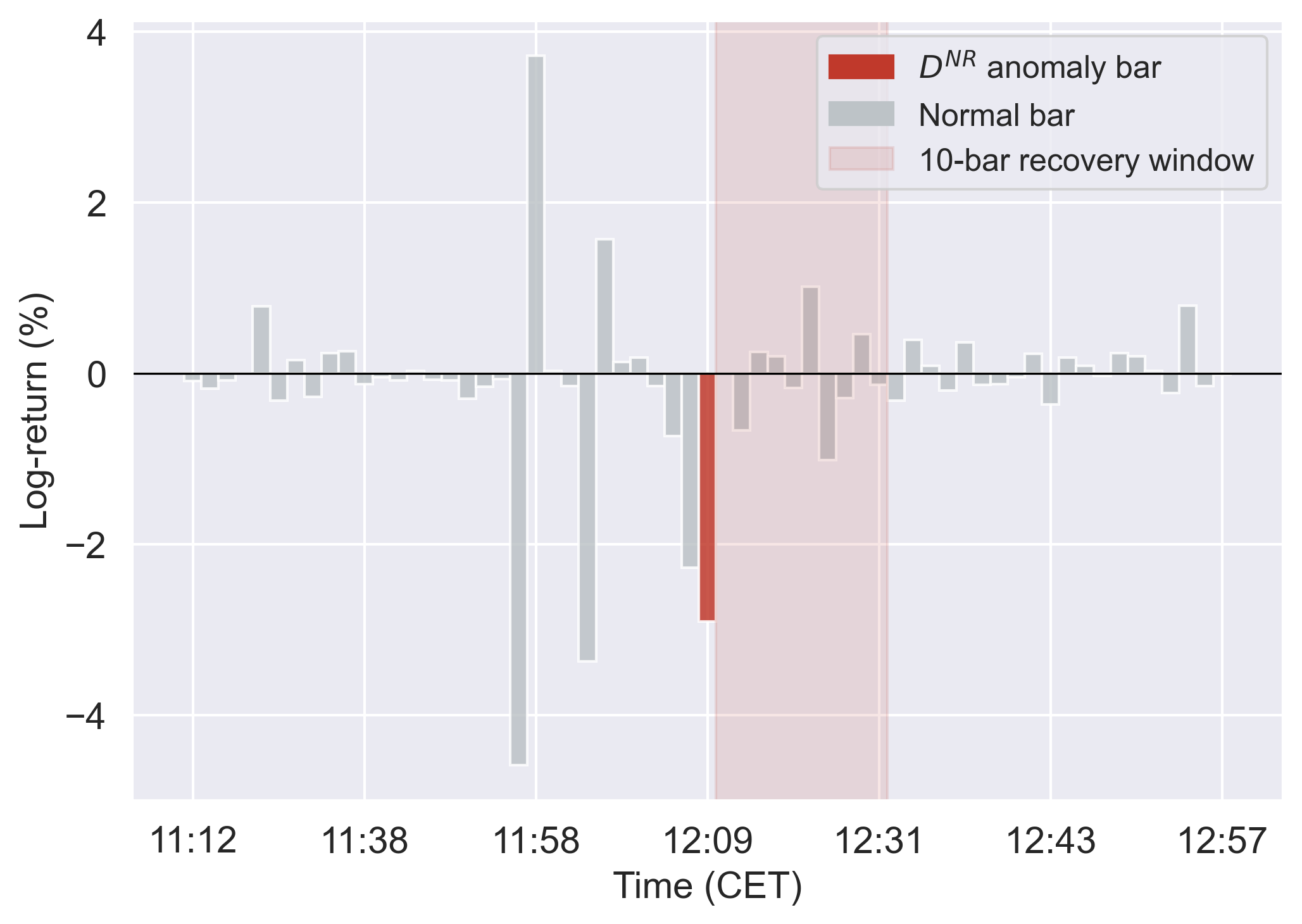}
        \caption{$D^{NR}$}
    \end{subfigure}
    \hfill
    \begin{subfigure}{.45\textwidth}
        \includegraphics[width=\linewidth]{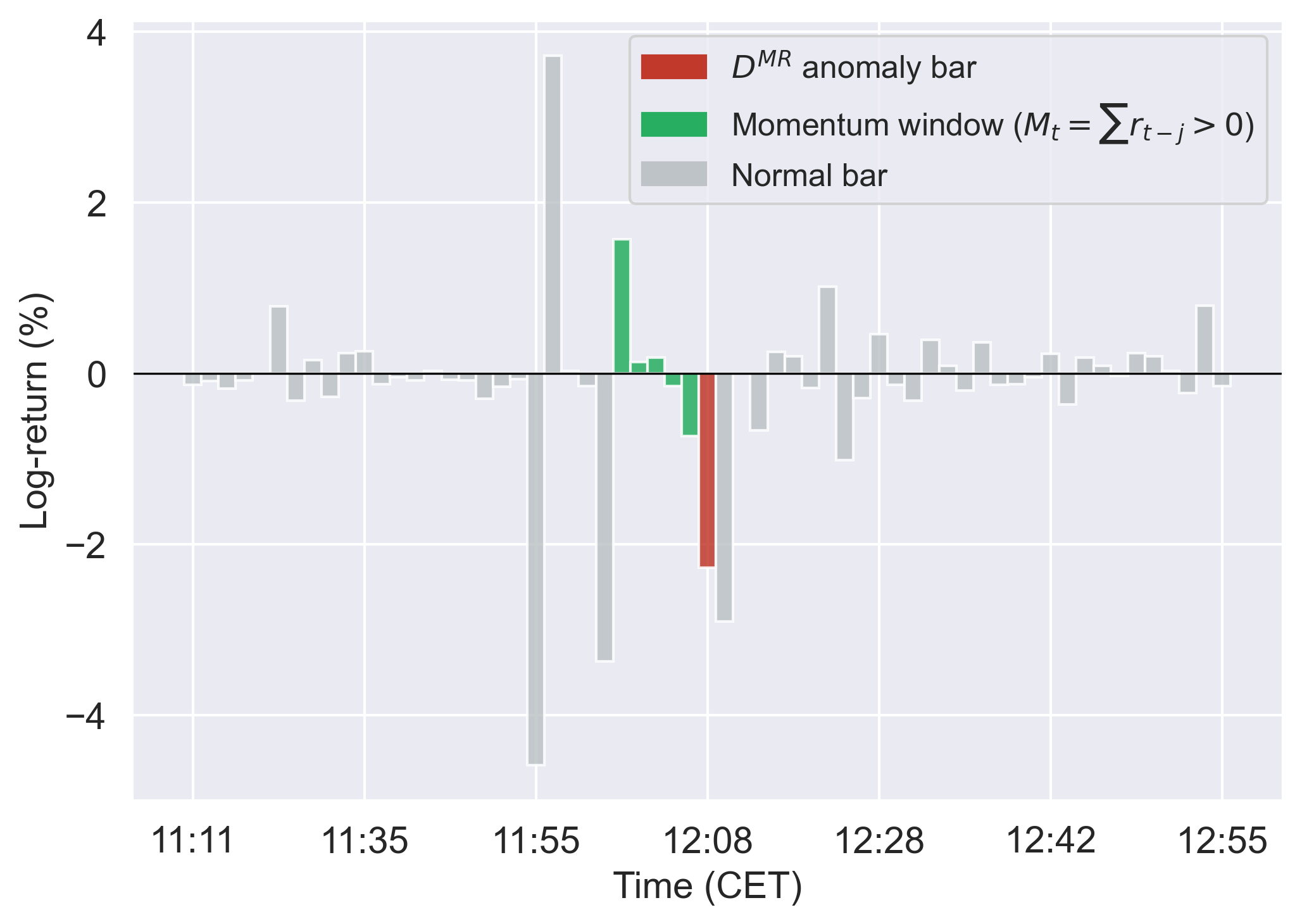}
        \caption{$D^{MR}$}
    \end{subfigure}
    \caption{Representative anomaly episodes for VBTC.XE. Anomaly bars are shown in red.}
    \label{fig:anomaly_episodes}
\end{figure}

In Figure~\ref{fig:anomaly_episodes} we illustrate example episodes for each of the four anomaly definitions, all for VBTC.XE. Panel~(a) shows a $D^{POT}$ episode in which several returns fall below the GPD threshold, with the largest decline reaching approximately $-4\%$. Panel~(b) shows a $D^{CV}$ episode in which the OLS residual $\hat{\varepsilon}_t$ exceeds the rolling 99th-percentile threshold, indicating a temporary pricing difference between VBTC.XE and VIRBTC.ST.
 Panel~(c) shows a $D^{NR}$ episode in which an extreme negative return is followed by a ten-bar observation window, shaded in red, with limited recovery by the end of the window. Panel~(d) illustrates a $D^{MR}$ episode in which the anomaly bar is preceded by five bars of positive cumulative momentum, highlighted in green.

\begin{table}[h]
\centering
\small
\caption{Number of anomalous bars identified by each definition over the full sample period, with the corresponding percentage of total valid bars in parentheses.}
\label{tab:event_counts}
\begin{tabular}{lrrrr}
\toprule
Definition & VBTC.XE & VETH.XE & VIRBTC.ST & VIRSETHS.ST \\
\midrule
POT    &  931 (0.599\%) &  475 (0.478\%) &  627 (0.542\%) & 279 (0.380\%) \\
Cross--venue &  795 (0.511\%) &  339 (0.341\%) &  795 (0.687\%) & 339 (0.462\%) \\
No recovery & 1{,}445 (0.929\%) &  826 (0.831\%) & 1{,}038 (0.897\%) & 581 (0.791\%) \\
Momentum reversal &  779 (0.501\%) &  410 (0.412\%) &  680 (0.588\%) & 368 (0.501\%) \\
\bottomrule
\end{tabular}
\end{table}

In Table~\ref{tab:event_counts} we report the number of anomalous bars identified 
by each definition over the full sample period. Each definition identifies anomalies in less than 1\% of valid bars.
VBTC.XE and VIRBTC.ST record the highest numbers of anomalies under most definitions because they have more valid bars than the Ethereum instruments.

The $D^{POT}$ definition identifies between 279 and 931 anomalous 
bars per instrument. The $D^{CV}$ definition gives the same number of anomalies for both instruments in each venue pair because the same bars are flagged for both. This results in 795 events for the Bitcoin pair (VBTC.XE and VIRBTC.ST) and 339 for the Ethereum pair (VETH.XE and VIRSETHS.ST).
The $D^{NR}$ definition produces the highest event rates among 
all four definitions, ranging from 0.791\% to 0.929\%. The $D^{MR}$ definition yields event 
rates comparable to $D^{POT}$, ranging from 0.412\% to 0.588\%.

\begin{table}[!ht]
\centering
\footnotesize
\caption{Median ratio (anomaly / normal) for selected microstructure
variables at anomaly bars, computed over all valid bars in the full
sample. For the non-negative variables, a ratio above one indicates
elevated values at anomaly bars.}

\label{tab:mwu_micro}
\setlength{\tabcolsep}{5pt}
\begin{tabular}{llcccc}
\toprule
Instrument & Definition
  & Eff.\ spread & Rel.\ volume & Amihud
  & Kyle $\lambda$  \\
\midrule
\multirow{4}{*}{VBTC.XE}
  & $D^{POT}$  & 2.05 & 4.95 &  1.59 &  3.42 \\
  & $D^{CV}$   & 1.45 & 2.25 &  0.74 &  1.27 \\
  & $D^{NR}$   & 1.16 & 4.98 &  1.25 &  2.74 \\
  & $D^{MR}$   & 1.22 & 3.55 &  1.75 &  3.17 \\
\midrule
\multirow{4}{*}{VETH.XE}
  & $D^{POT}$  & 2.41 & 4.98 &  2.38 &  4.47 \\
  & $D^{CV}$   & 1.21 & 2.71 &  0.62 &  1.23 \\
  & $D^{NR}$   & 1.22 & 5.08 &  1.29 &  3.02 \\
  & $D^{MR}$   & 1.42 & 3.75 &  2.61 &  4.21  \\
\midrule
\multirow{4}{*}{VIRBTC.ST}
  & $D^{POT}$  & 3.90 & 2.44 &  6.55 &  8.15 \\
  & $D^{CV}$   & 1.81 & 1.59 &  6.13 &  7.15 \\
  & $D^{NR}$   & 1.37 & 2.67 &  3.50 &  5.06 \\
  & $D^{MR}$   & 1.58 & 2.12 &  4.80 &  5.91  \\
\midrule
\multirow{4}{*}{VIRSETHS.ST}
  & $D^{POT}$  & 3.99 & 2.58 &  8.57 & 11.24 \\
  & $D^{CV}$   & 1.71 & 2.20 &  4.72 &  6.30 \\
  & $D^{NR}$   & 1.36 & 2.78 &  3.39 &  5.07  \\
  & $D^{MR}$   & 1.44 & 2.16 &  4.60 &  5.89  \\
\bottomrule
\end{tabular}
\end{table}

After examining how often each anomaly type occurs, we next test whether anomaly bars are associated with different microstructure conditions than normal trading bars.
 For each anomaly definition and each instrument, we split all valid bars into anomaly and normal bars and compare the distributions of the microstructure variables between the two groups using the two-sided Mann--Whitney U test. The tests reject the null hypothesis of equal distributions for all reported variable--definition--instrument combinations. 
 As an interpretable effect-size measure, Table~\ref{tab:mwu_micro} reports the median ratio of selected 
microstructure variables at anomaly bars relative to normal bars. Let 
$v_t$ denote the trade volume in units, $V_t$ the trading value in EUR, 
$p_t$ the transaction price, and $m_t = (\text{ask}_t + \text{bid}_t)/2$ 
the prevailing mid-quote at bar $t$. The effective spread is calculated as $\text{ES}_t = 2|p_t-m_t|/m_t$, where $p_t$ is the transaction price and $m_t$ is the mid-quote. It measures the estimated round-trip trading cost relative to the mid-quote \citep{glosten1985bid}. Relative volume is $\text{RV}_t = v_t / \bar{v}_t$, where $\bar{v}_t$ is the 
rolling 60-bar mean trade volume, capturing abnormal trading activity 
at bar $t$. The relative spread is defined as
\begin{equation}
\text{RS}^{spr}_t =
\frac{\text{ES}_t}{\overline{\text{ES}}_{60,t}},
\end{equation}
where $\overline{\text{ES}}*{60,t}$ denotes the mean effective spread over the preceding 60 bars. The relative spread measures the effective spread at bar $t$ relative to its recent level. It is included in the prediction analysis but is not reported in Table~\ref{tab:mwu_micro}.
 Following \citet{amihud2002illiquidity}, who scale dollar volume by $10^6$ (i.e.\ express it in millions) to avoid extremely small ratio values, the Amihud illiquidity ratio is defined at the one-minute level as $\text{AMIH}_t = |r_t| / (V_t \cdot 10^{-6})$, where $V_t$ is expressed in EUR. Kyle’s $\lambda$ \citep{kyle1985continuous} is approximated as $\lambda_t = |r_t|/\sqrt{v_t}$ and measures the price change relative to trading volume.
The order imbalance is defined as
\begin{equation}
\text{OIB}_t = \frac{v_t^{a} - v_t^{b}}{v_t},
\end{equation}
where $v_t^{a}$ and $v_t^{b}$ denote the ask-side and bid-side executed volume, respectively. 
For the non-negative variables reported in Table~\ref{tab:mwu_micro}, we calculate the ratio $\tilde{X}^{A}/\tilde{X}^{N}$, where $\tilde{X}^{A}$ is the median for anomaly bars and $\tilde{X}^{N}$ is the median for normal bars.

The results differ across variables and trading venues. The $D^{POT}$ bars have the highest effective-spread and Kyle's $\lambda$ ratios for all four instruments. For VBTC.XE, these ratios are 2.05 and 3.42, respectively. The corresponding values are particularly high for VIRSETHS.ST, reaching 3.99 for the effective spread and 11.24 for Kyle's $\lambda$. Relative volume, however, is highest for $D^{NR}$ for all four instruments, ranging from 2.67 to 5.08 times the normal level. For the Xetra instruments, the highest Amihud ratios are observed for $D^{MR}$, while for the Stockholm instruments they are highest for $D^{POT}$.

The relation between $D^{CV}$, $D^{NR}$, and $D^{MR}$ differs across venues. For the Xetra instruments, $D^{CV}$ has lower Amihud and Kyle's $\lambda$ ratios than $D^{NR}$ and $D^{MR}$. For the Stockholm instruments, $D^{CV}$ has higher Amihud and Kyle's $\lambda$ ratios than $D^{NR}$ and $D^{MR}$, although these values remain below those observed for $D^{POT}$. The effective-spread ratios for $D^{NR}$ and $D^{MR}$ are more moderate for the Xetra instruments, ranging from 1.16 to 1.42.

OIB provides additional information about the direction of trading activity during anomaly bars. For $D^{POT}$, $D^{NR}$, and $D^{MR}$, the median order imbalance changes from positive values during normal trading to negative values during anomaly bars. The corresponding anomaly-to-normal median ratios are negative for all four instruments under each of these definitions. This suggests that these anomalies are associated with stronger selling pressure. By contrast, the directional difference is weaker and less uniform for $D^{CV}$, with ratios of 0.25 for VBTC.XE, 0.47 for VETH.XE, 0.12 for VIRBTC.ST, and $-0.04$ for VIRSETHS.ST.

\begin{figure}[!ht]
    \begin{subfigure}{.5\textwidth}
        \includegraphics[width=0.92\linewidth]{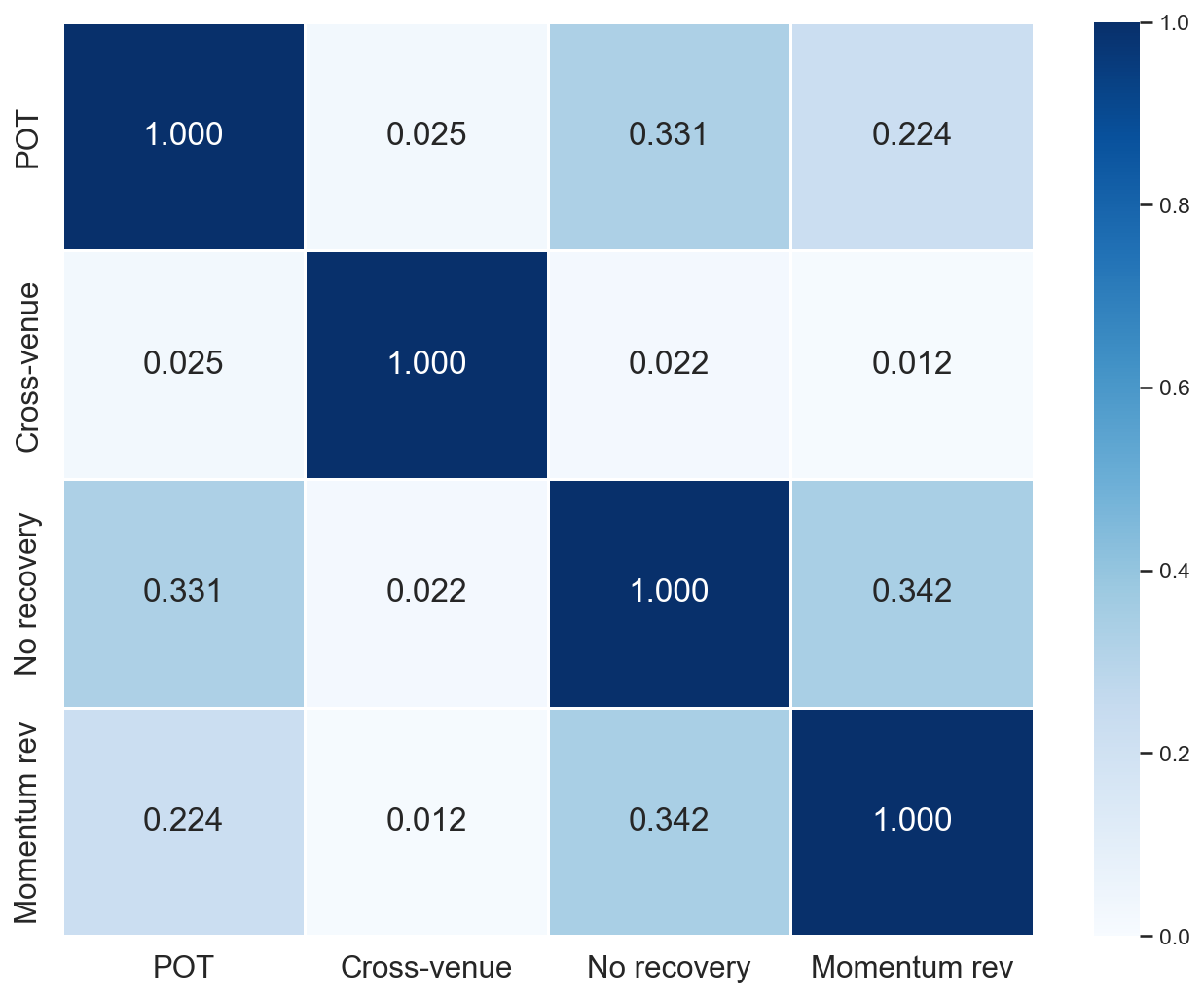}
        \caption{VBTC.XE}
    \end{subfigure}
    \begin{subfigure}{.5\textwidth}
        \includegraphics[width=0.92\linewidth]{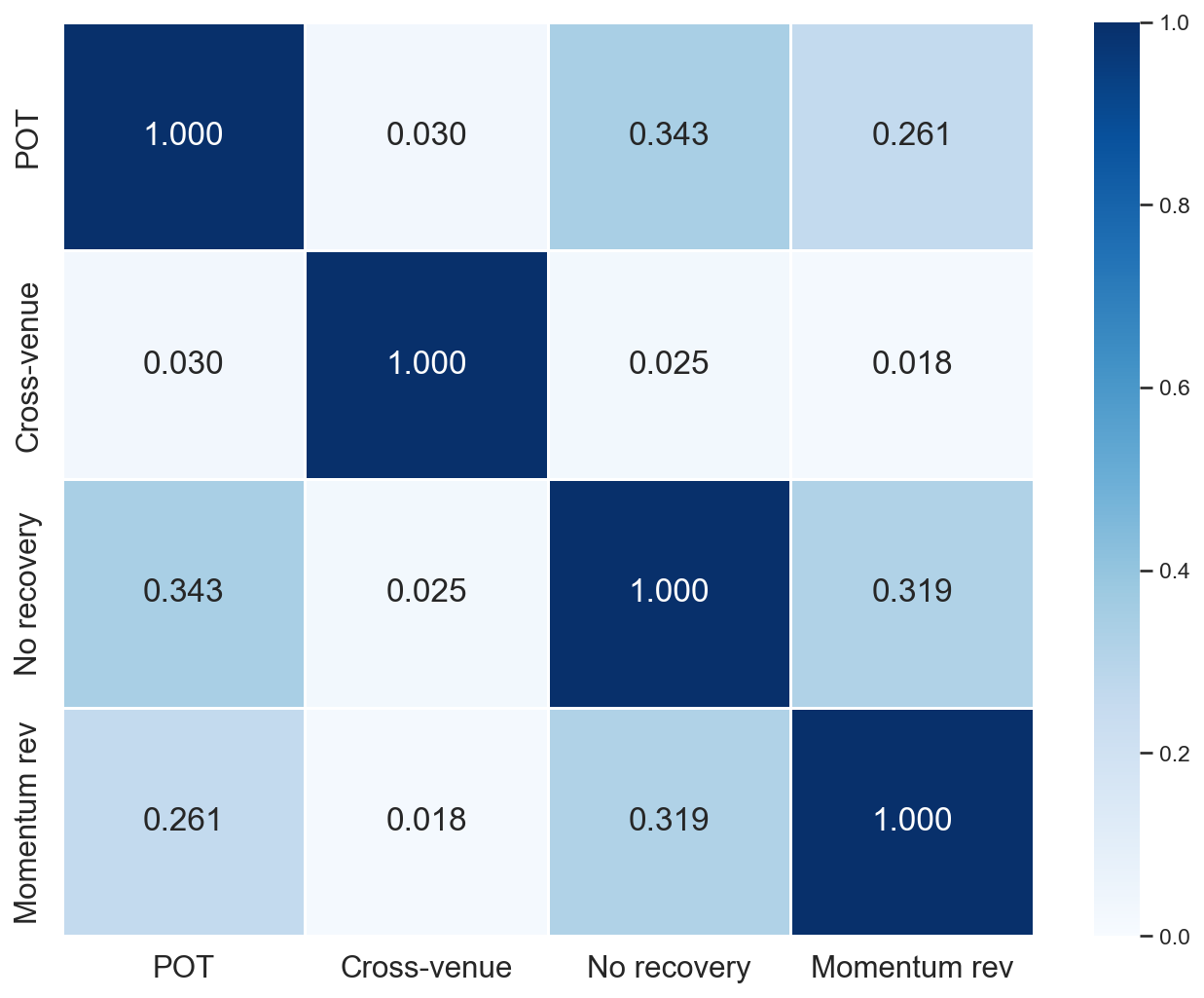}
        \caption{VETH.XE}
    \end{subfigure}
    \begin{subfigure}{.5\textwidth}
        \includegraphics[width=0.92\linewidth]{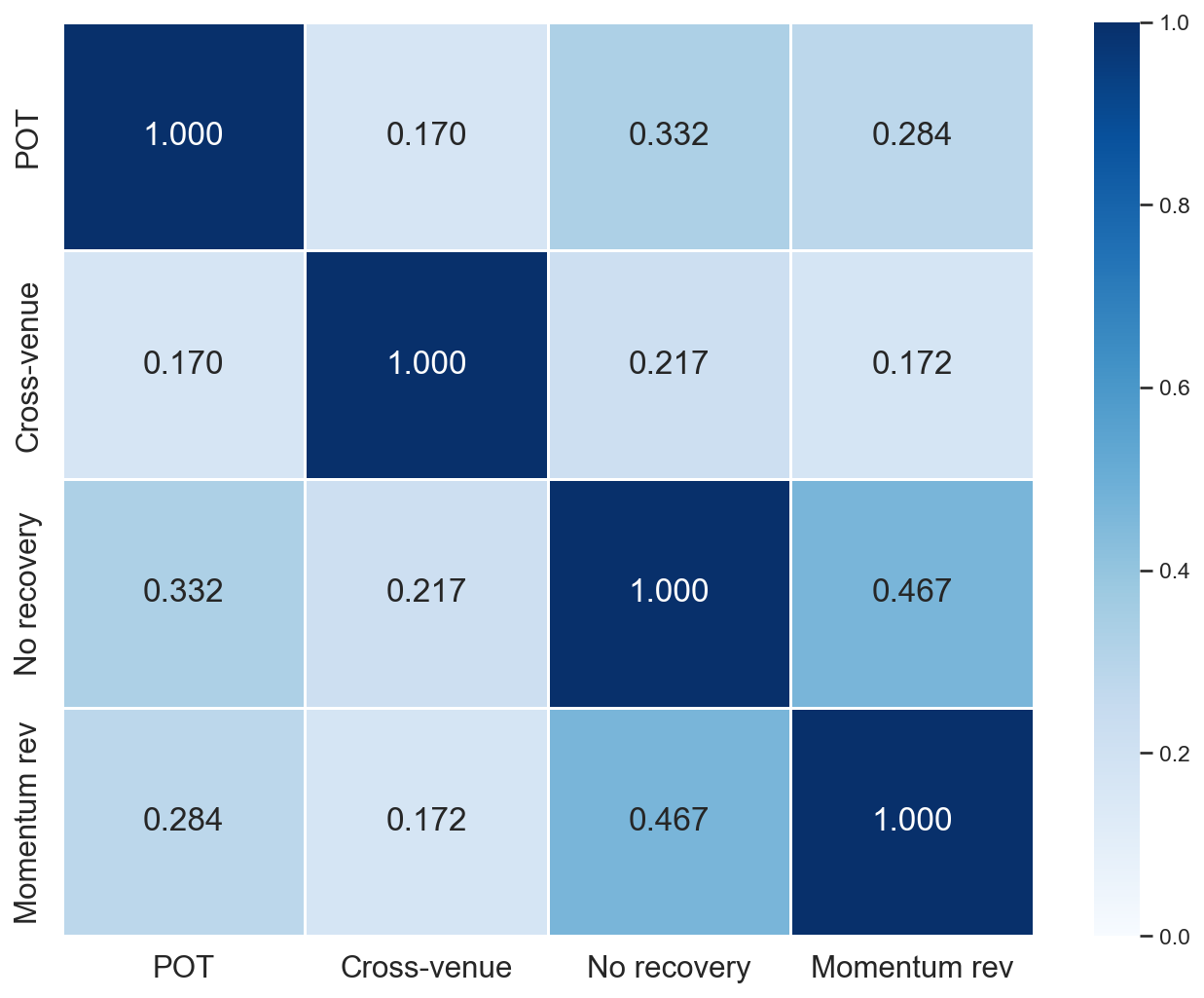}
        \caption{VIRBTC.ST}
    \end{subfigure}
    \begin{subfigure}{.5\textwidth}
        \includegraphics[width=0.92\linewidth]{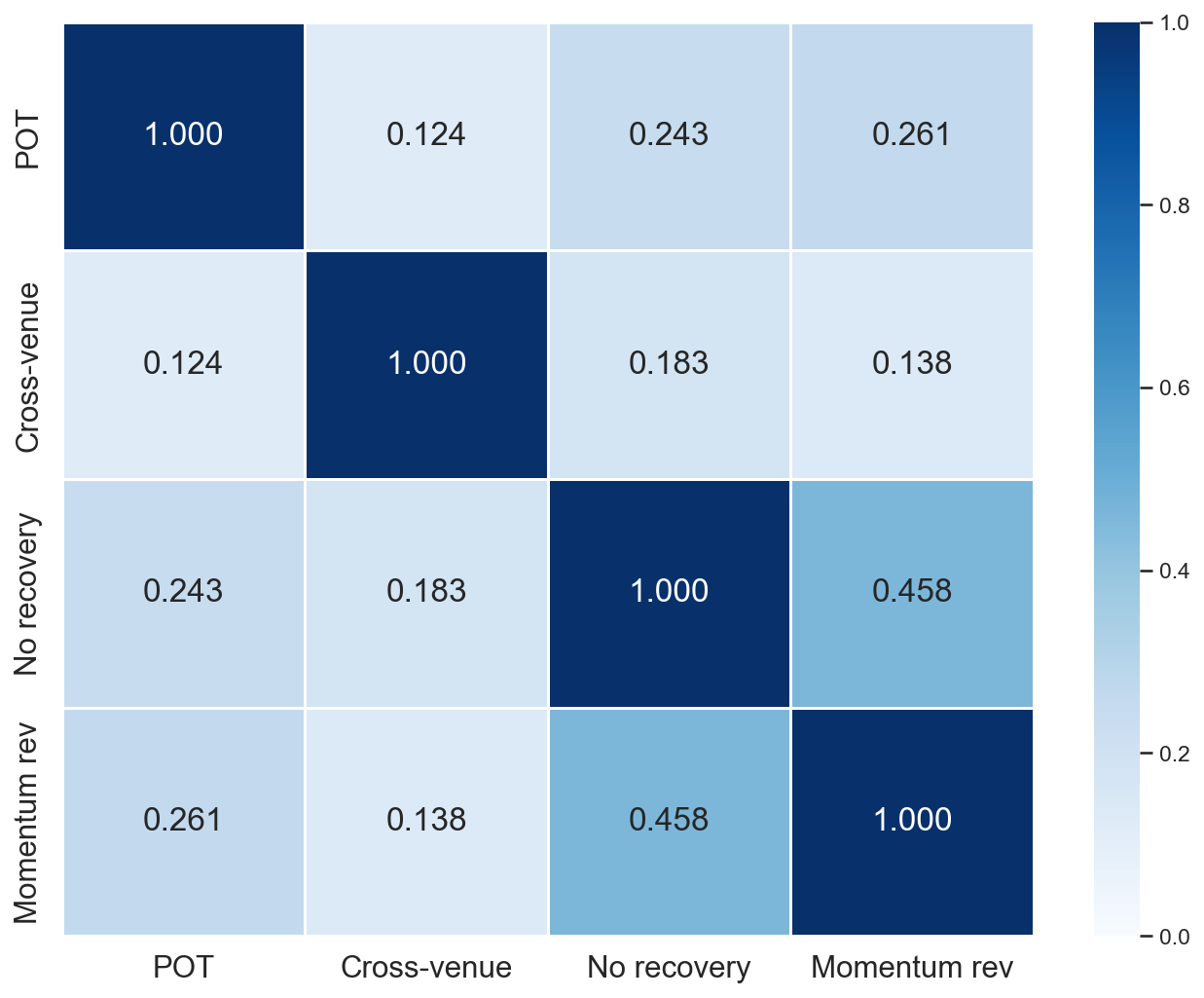}
        \caption{VIRSETHS.ST}
    \end{subfigure}
    \caption{Pairwise Jaccard similarity between anomaly definitions for each instrument over the full sample period.}
    \label{fig:jaccard}
\end{figure}

This raises the question of whether the four definitions identify the same or different types of anomalous behaviour.
Figure~\ref{fig:jaccard} shows the pairwise Jaccard similarity between the anomaly definitions for each instrument \citep{real1996probabilistic}. The Jaccard index measures the overlap between two anomaly sets as the number of bars flagged by both definitions divided by the number flagged by at least one of them.
 It ranges from 0 to 1, where 0 indicates no overlap and 1 indicates complete agreement.
Across all four instruments, $D^{CV}$ has the lowest overlap with the other anomaly definitions, with Jaccard values between 0.012 and 0.030.
 It captures venue-specific pricing differences that rarely occur at the same time as the extreme return events identified by the other definitions.

The highest pairwise similarity is observed between $D^{NR}$ 
and $D^{MR}$, with Jaccard values from 0.319 to 0.467. This is because both definitions identify returns below the same rolling first-percentile threshold and differ only in the additional condition applied to the surrounding price movements.
The similarity between $D^{POT}$ and $D^{NR}$ is moderate, ranging from 
0.243 to 0.343, while the overlap between $D^{POT}$ and $D^{MR}$ is 
somewhat lower at 0.224--0.284. These values confirm that the novel 
definitions overlap with a subset of $D^{POT}$ events while 
also identifying events not flagged by the GPD threshold alone. Overall, the low to moderate Jaccard values show that the four definitions capture different types of intraday events. They therefore do not simply identify the same observations under different labels, but provide complementary views of abnormal behaviour.

\begin{figure}[p]
    \begin{subfigure}{\textwidth}
        \centering
        \includegraphics[width=0.68\linewidth]{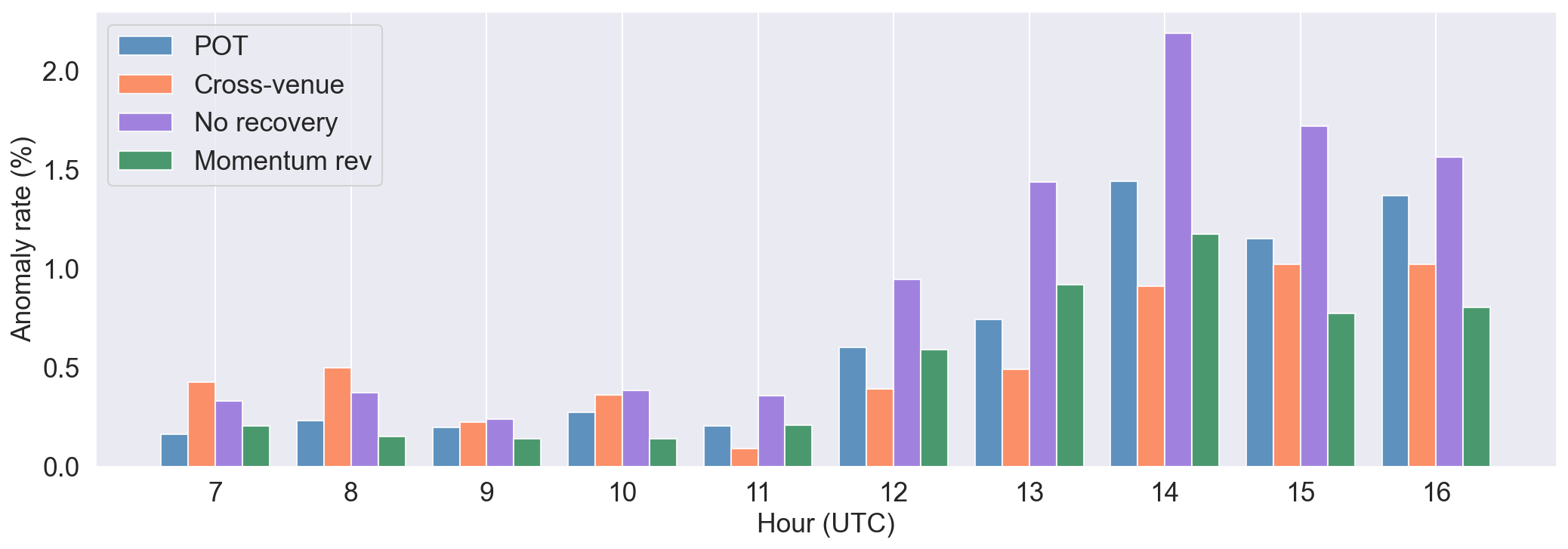}
        \caption{VBTC.XE}
    \end{subfigure}
    \begin{subfigure}{\textwidth}
        \centering
        \includegraphics[width=0.68\linewidth]{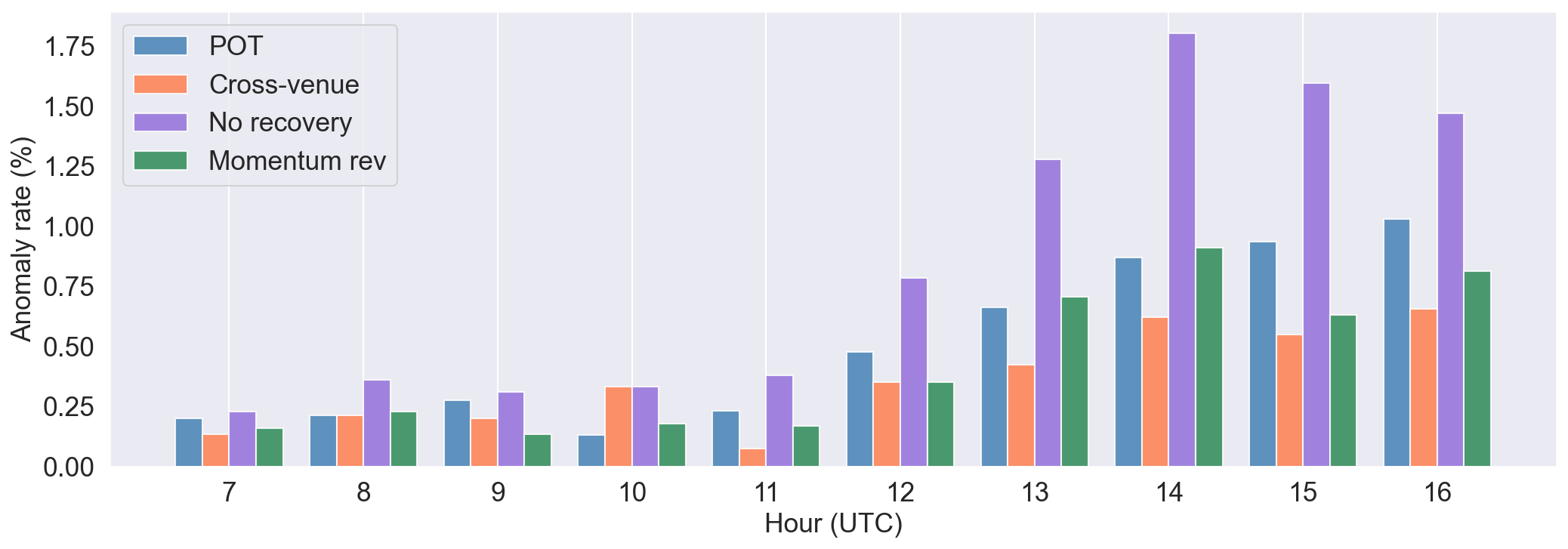}
        \caption{VETH.XE}
    \end{subfigure}
    \begin{subfigure}{\textwidth}
        \centering
        \includegraphics[width=0.68\linewidth]{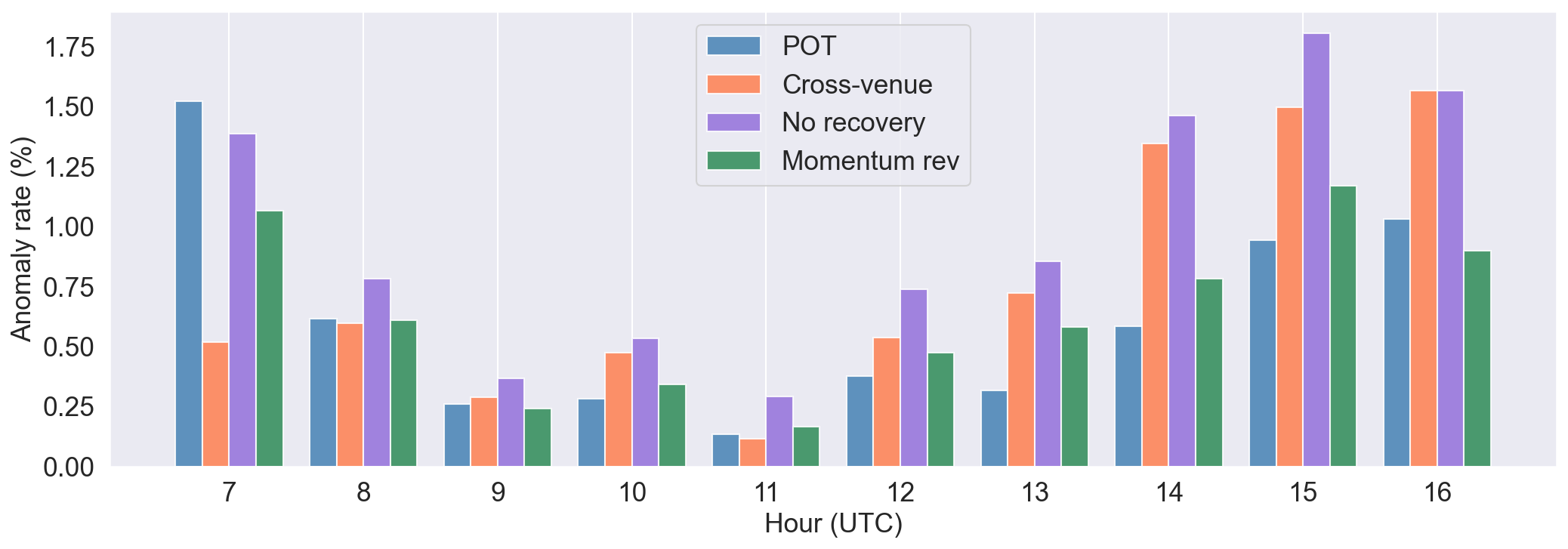}
        \caption{VIRBTC.ST}
    \end{subfigure}
    \begin{subfigure}{\textwidth}
        \centering
        \includegraphics[width=0.68\linewidth]{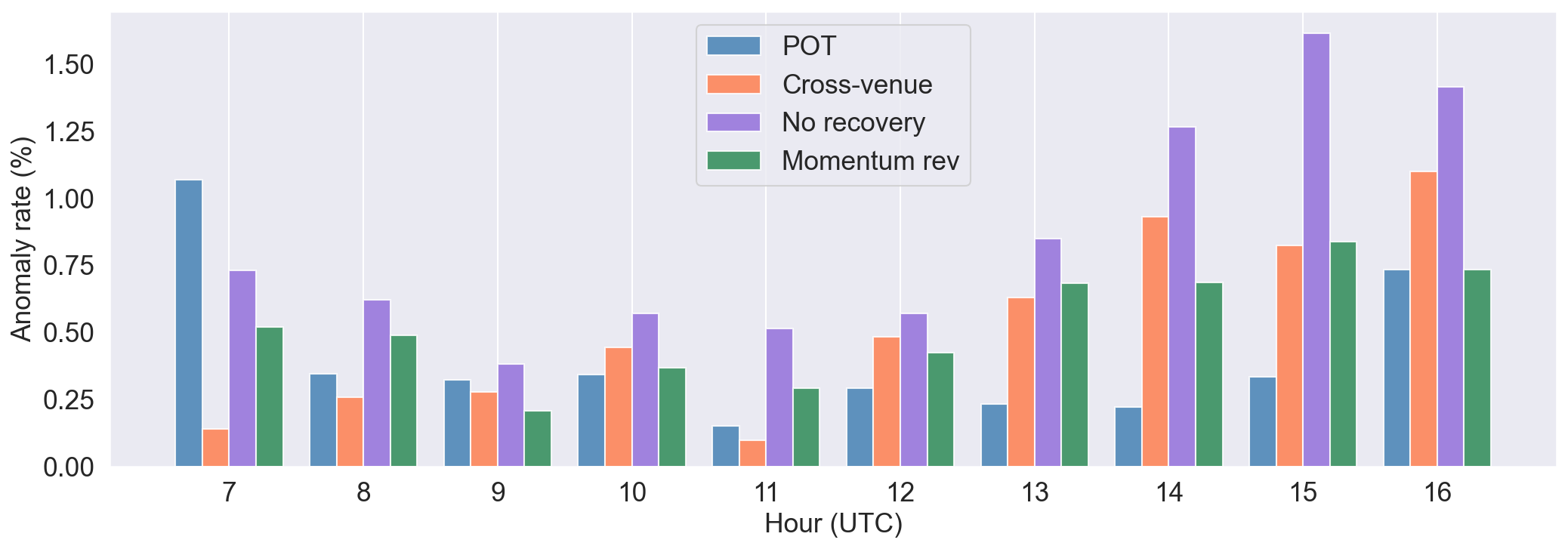}
        \caption{VIRSETHS.ST}
    \end{subfigure}
    \caption{Hourly anomaly rate (\% of bars flagged as anomalous) during exchange hours, averaged over the full sample.}
    \label{fig:patterns_hourly}
\end{figure}

Beyond their overlap with one another, the four definitions may also differ in when during the trading day or week they tend to occur.
Figure~\ref{fig:patterns_hourly} presents the hourly anomaly rate for 
each definition during exchange hours, averaged over the full sample 
period. The intraday patterns differ markedly across definitions and 
venues, and are broadly consistent with the intraday volatility and 
volume patterns documented in Section~\ref{sec:data}.

For the Xetra instruments, all definitions show a pronounced increase 
in anomaly rates towards the end of the trading session, with rates 
peaking in the 15:00--16:00 interval. This is evident 
for $D^{NR}$, which reaches anomaly rates of approximately 2.0\% for 
VBTC.XE and 1.75\% for VETH.XE in the final hours of trading, 
consistent with the elevated volatility observed during this period. 
The morning hours are characterised by low and 
relatively stable anomaly rates across all definitions.

The Stockholm instruments display a different intraday pattern, with 
elevated anomaly rates at the open followed by a 
decline through the mid-session. This mirrors the opening volatility 
spike documented in Figure~\ref{fig:intraday_patterns} and reflects 
the rapid resolution of overnight information at the start of the 
Nasdaq Stockholm session. A secondary increase in anomaly rates is 
observed in the afternoon, consistent with the influence of US market 
opening hours on cryptocurrency prices.

Across all instruments, $D^{CV}$ produces the lowest 
intraday anomaly rates, while $D^{NR}$ produces the highest, 
particularly in the afternoon hours. The intraday patterns of $D^{POT}$ 
and $D^{MR}$ are broadly similar to each other, reflecting their common 
dependence on the rolling return threshold. Anomaly rates vary throughout the trading day, with unusual price movements occurring more often near the market open and close.

\begin{figure}[h]
    \begin{subfigure}{.5\textwidth}
        \includegraphics[width=\linewidth]{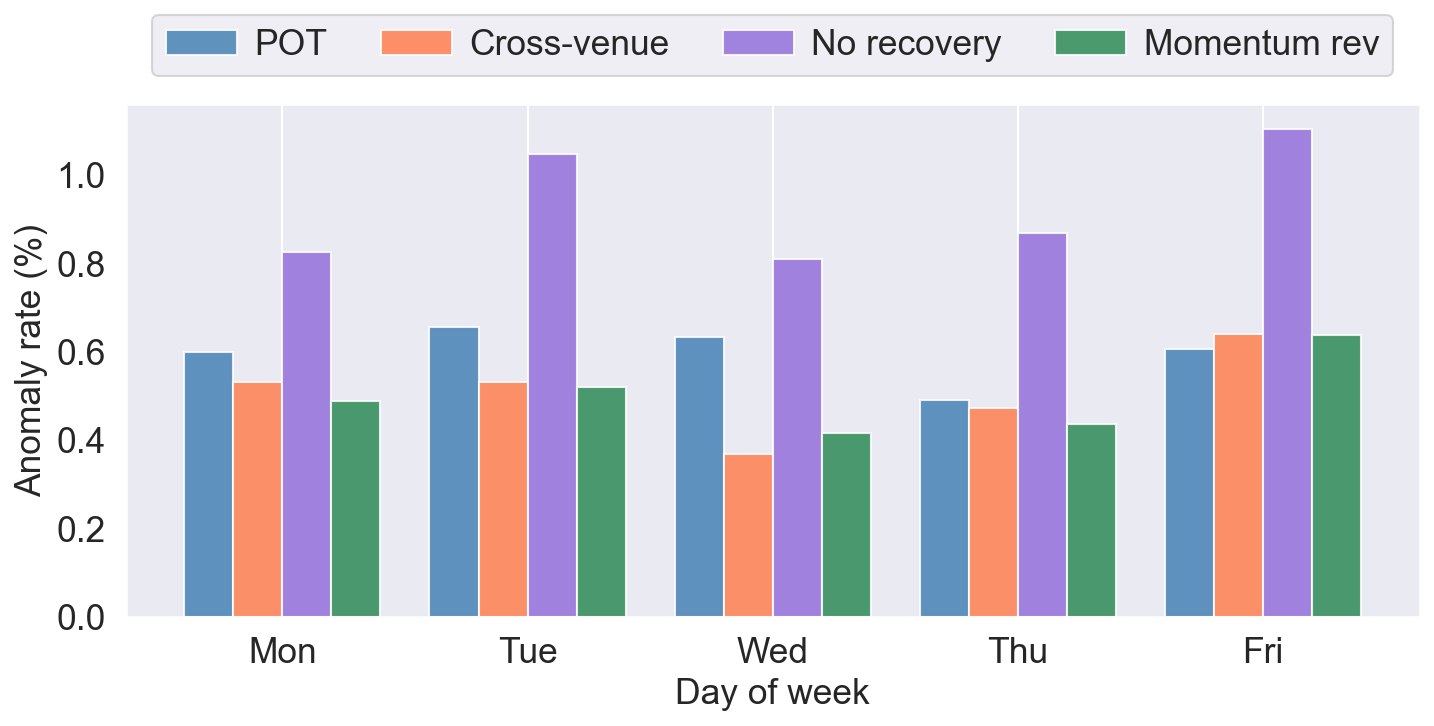}
        \caption{VBTC.XE}
    \end{subfigure}
    \begin{subfigure}{.5\textwidth}
        \includegraphics[width=\linewidth]{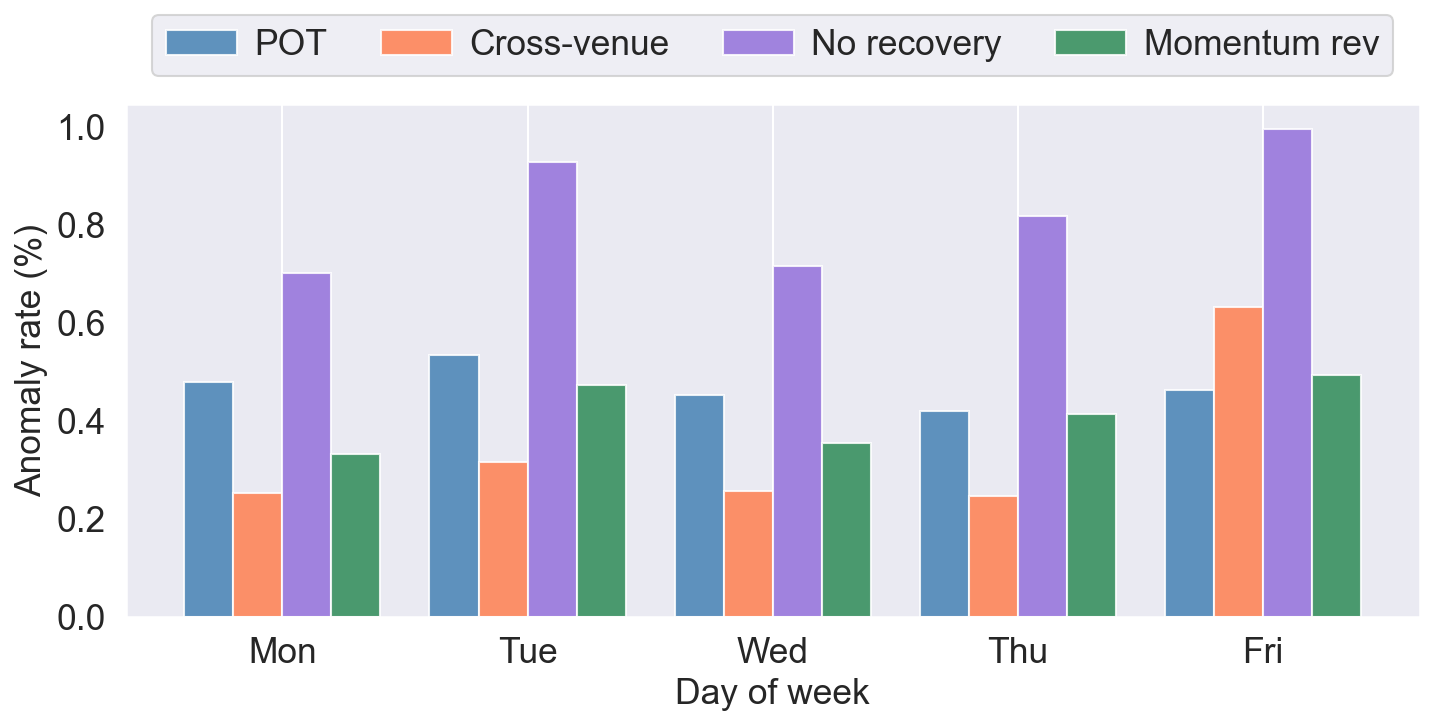}
        \caption{VETH.XE}
    \end{subfigure}
    \begin{subfigure}{.5\textwidth}
        \includegraphics[width=\linewidth]{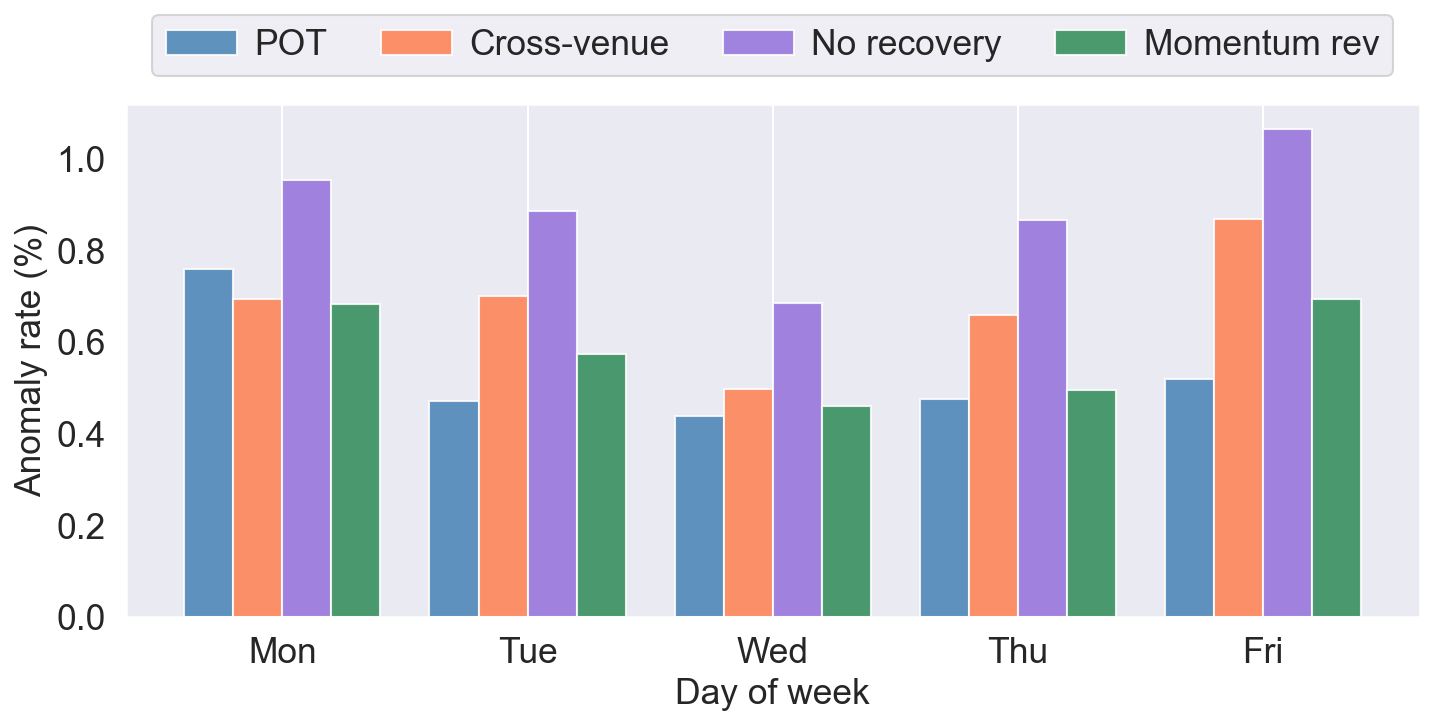}
        \caption{VIRBTC.ST}
    \end{subfigure}
    \begin{subfigure}{.5\textwidth}
        \includegraphics[width=\linewidth]{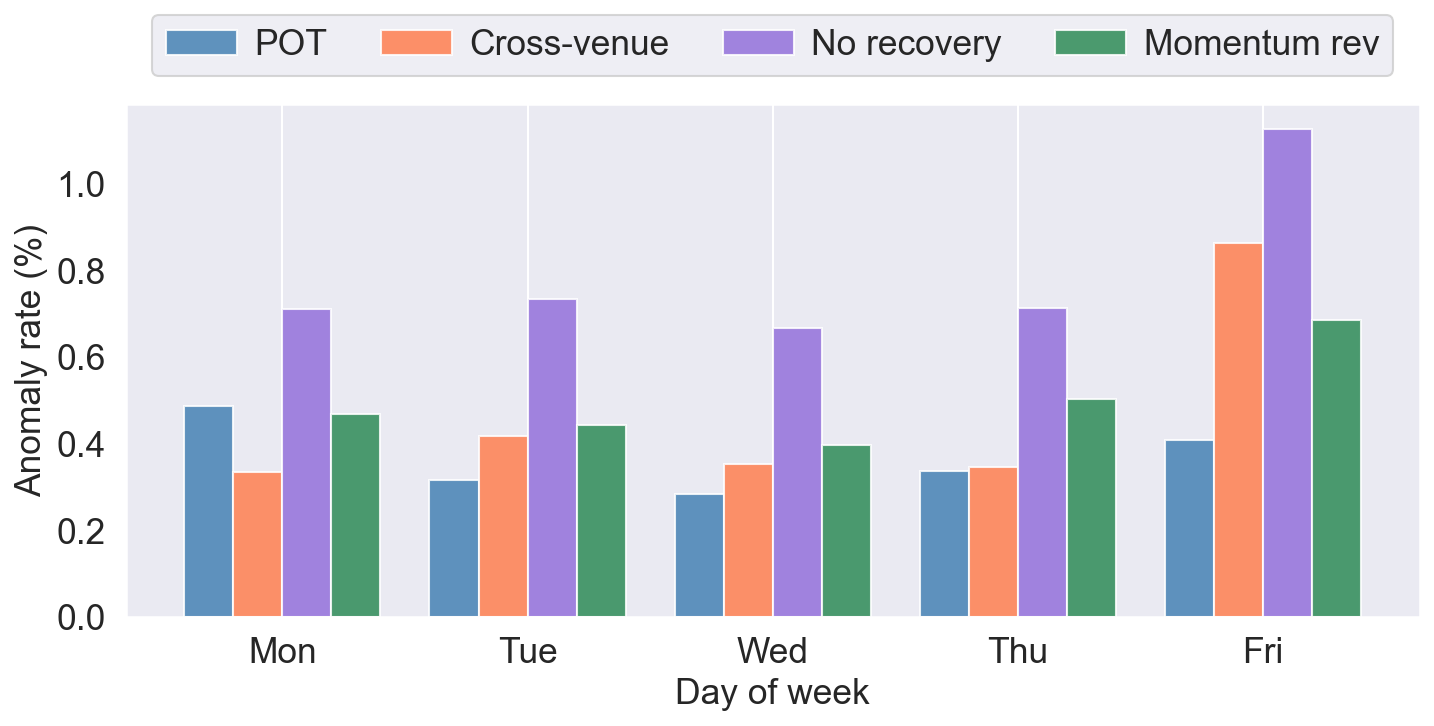}
        \caption{VIRSETHS.ST}
    \end{subfigure}
    \caption{Day-of-week anomaly rate (\% of bars flagged as anomalous) for Monday through Friday, averaged over the full sample period January 2024 -- December 2025.}
    \label{fig:patterns_dow}
\end{figure}

In Figure~\ref{fig:patterns_dow} we show the average day-of-week anomaly rates for each definition over the full sample period. The patterns are generally similar across instruments, with two main features standing out.
First, anomaly rates are higher on Fridays for most definitions 
and all four instruments. For the no-recovery definition $D^{NR}$, 
Friday rates reach approximately 1.0\% for most instruments, 
substantially above the mid-week levels. This is consistent with 
reduced liquidity provision towards the end of the trading week and 
the accumulation of uncertainty ahead of the weekend, during which 
cryptocurrency markets continue to trade while European ETP markets 
are closed.

Second, anomaly rates are also higher on Tuesdays for several instruments and definitions, especially for $D^{NR}$ and $D^{CV}$. Anomaly rates are lowest on Wednesdays
across all definitions and instruments, suggesting that mid-week 
trading conditions are the most stable.
Among the four definitions, $D^{NR}$ varies the most across days of the week.
 The profiles of $D^{CV}$ and $D^{MR}$ are more stable across the week.
The day-of-week pattern is similar on both exchanges, suggesting that it reflects the broader cryptocurrency market rather than factors specific to either venue.

\section{Anomaly prediction}
\label{sec:prediction}

After analysing the statistical properties and microstructure patterns of the four anomaly definitions, we examine whether anomalous bars can be predicted. For each definition, we test if the next bar can be classified as anomalous using information available at the current time.
We consider five prediction targets: the four individual anomaly definitions, $D^{POT}$, $D^{CV}$, $D^{NR}$, and $D^{MR}$, and the composite label
$D^{ANY}_t = D^{POT}_t \vee D^{CV}_t \vee D^{NR}_t \vee D^{MR}_t$.
The composite label equals one when at least one of the four anomaly definitions is triggered. It may be useful in practice because it could be applied in systems that monitor several types of risk at the same time.

We evaluate four classifiers that are well-established in the
financial machine learning literature: logistic regression, random forest, extreme gradient boosting, and 
light gradient boosting machine. 
All models are trained on the dataset
$\mathcal{D}=\{(x_t,y_{t+1})\}_{t=1}^{N-1}$,
where $x_t\in\mathbb{R}^p$ is the vector of features available at bar $t$ and $y_{t+1}\in\{0,1\}$ is the binary anomaly label for the next bar. The goal is to learn a function
$f:\mathbb{R}^p\to[0,1]$ that estimates the probability
$P(y_{t+1}=1\mid x_t)$ on unseen data. All classifiers use class weights to reduce the effect of the strong class imbalance caused by the small number of anomaly events. The same hyperparameters are used for all instruments and prediction targets rather than being tuned separately for each model. This reduces the risk of overfitting.

LR estimates the probability of an anomaly from a weighted combination of the input features using a sigmoid function \citep{hosmer2013applied}.
 The predicted probability
is given by
\begin{equation}
\hat{P}(y_{t+1} = 1 \mid x_t)
= \sigma(\beta_0 + \beta^\top x_t)
= \frac{1}{1 + \exp(-\beta_0 - \beta^\top x_t)},
\end{equation}
where $\beta_0 \in \mathbb{R}$ is the intercept and
$\beta \in \mathbb{R}^p$ is the vector of coefficients.
The parameters are estimated by maximising the
regularised log-likelihood
\begin{equation}
\ell(\beta_0, \beta) =
\sum_{t=1}^{N-1}
\left[
y_{t+1}\log\hat{P}_t+
(1-y_{t+1})\log(1-\hat{P}_t)
\right] - \lambda \|\beta\|_2^2,
\end{equation}
where $\lambda > 0$ controls the strength of $L_2$
regularisation.

To capture non-linear relationships between the features and anomaly occurrence, we also use RF, which combines multiple decision trees \citep{breiman2001random}.
Each tree is trained on a bootstrap sample, and only a random subset of features is considered at each split. This reduces the correlation between trees.
The algorithm builds $B$ classification trees
$\{f_b(x)\}_{b=1}^B$, and the predicted probability is
obtained by averaging their class probability outputs
\begin{equation}
\hat{P}(y = 1 \mid x) =
\frac{1}{B} \sum_{b=1}^B f_b(x),
\end{equation}
where $f_b(x) \in [0,1]$ denotes the class probability
estimate of the $b$-th tree.

We also use two gradient boosting methods. These models build decision trees one after another, with each new tree reducing the errors made by the previous trees. LightGBM is a gradient boosting method based on decision trees \citep{wang2022corporat}.
It improves standard gradient boosting by using histogram-based splits and growing trees leaf by leaf.
LightGBM constructs an additive model of trees
\begin{equation}
F_m(x) = \sum_{k=1}^m \gamma_k f_k(x),
\end{equation}
where $f_k(x)$ is the $k$-th tree and $\gamma_k$ is its
weight. At iteration $m$, the objective is minimised using
a second-order Taylor expansion of the loss $L(y, F(x))$
\begin{equation}
L^{(m)} \approx \sum_{i=1}^N \left[
g_{im} f_m(x_i) +
\tfrac{1}{2} h_{im} f_m(x_i)^2
\right],
\end{equation}
where
\begin{equation}
g_{im} = \frac{\partial L(y_i, F(x_i))}{\partial F(x_i)},
\qquad
h_{im} = \frac{\partial^2 L(y_i, F(x_i))}{\partial F(x_i)^2}
\end{equation}
are the first and second derivatives of the loss. The model
is updated as
$F_m(x) = F_{m-1}(x) + \nu \gamma_m f_m(x)$,
where $\nu \in (0,1]$ is the learning rate.

XGBoost is a gradient boosting method that uses regularisation to reduce the risk of overfitting \citep{ramraj2016experimenting}.
XGBoost constructs an additive model of trees
\begin{equation}
F_m(x) = \sum_{k=1}^m f_k(x).
\end{equation}
The training objective at iteration $m$ is
\begin{equation}
\mathcal{L}^{(m)} =
\sum_{i=1}^N L(y_i, F_{m-1}(x_i) + f_m(x_i))
+ \Omega(f_m),
\end{equation}
where $\Omega(f_m) = \gamma J + \frac{1}{2}\lambda
\sum_{j=1}^J w_j^2$ penalises model complexity through
the number of leaves $J$ and leaf weights $w_j$.
Using a second-order Taylor expansion
\begin{equation}
\mathcal{L}^{(m)} \approx
\sum_{i=1}^N \left[
g_{im} f_m(x_i) +
\tfrac{1}{2} h_{im} f_m(x_i)^2
\right] + \Omega(f_m).
\end{equation}
The optimal weight for each leaf $j$ is
\begin{equation}
w_j^* = -\frac{\sum_{i \in I_j} g_{im}}
              {\sum_{i \in I_j} h_{im} + \lambda},
\end{equation}
where $I_j$ denotes the set of observations assigned to leaf $j$,
and the corresponding optimal objective value is
\begin{equation}
\mathcal{L}^{(m)}_{\mathrm{opt}} =
-\frac{1}{2} \sum_{j=1}^J
\frac{\left(\sum_{i \in I_j} g_{im}\right)^2}
     {\sum_{i \in I_j} h_{im} + \lambda}
+ \gamma J.
\end{equation}

All models are evaluated out of sample using a chronological split. The classifiers are trained once on the first 80\% of the data and tested on the remaining 20\%, without retraining.
We consider two ways of constructing the features. In the session-based approach, rolling features are calculated using data from the current trading session only. Predictions start from the 30th bar of each session, ensuring that the shortest rolling windows contain enough same-day observations. This approach represents an intraday monitoring system that restarts at the beginning of each trading day and does not use information from the overnight period.

In the cumulative approach, rolling features use the full sequence of active bars, including windows that extend across trading sessions.
 This approach uses historical information from previous trading sessions.
In both approaches, the features use only information available before bar $t$. However, for the $D^{NR}$ target, the label depends on the $K$ bars following the predicted bar. As a result, the features are causal, but the final label can only be observed after a delay of $K$ bars.

For each bar $t$, we create a vector of $p = 34$ features using price and microstructure information available before that bar. All features are shifted by one bar to avoid look-ahead bias. The features are divided into two groups: price and volatility features, and microstructure features.

We compute six features over rolling windows of
$w \in \{5, 15, 30\}$ bars \citep{nti2020systematic, tsay2005analysis}.
The rolling standard deviation and mean of returns are
\begin{equation}
\text{std}_w(t) = \sqrt{\frac{1}{w}\sum_{j=1}^{w}
\bigl(r_{t-j} - \bar{r}_{w,t}\bigr)^2},
\qquad
\bar{r}_{w,t} = \frac{1}{w}\sum_{j=1}^{w} r_{t-j},
\end{equation}
where $r_{t-j}$ denotes the one-minute log-return at bar $t-j$
and $\bar{r}_{w,t}$ is the rolling mean over the preceding
$w$ bars, capturing short-term volatility and price drift.
The rolling skewness
\begin{equation}
\text{skew}_w(t) =
\frac{\frac{1}{w}\sum_{j=1}^{w}(r_{t-j} - \bar{r}_{w,t})^3}
     {\text{std}_w(t)^3},
\end{equation}
captures asymmetry in the recent return distribution,
where $\text{std}_w(t)^3$ is the cube of the rolling standard
deviation serving as a normalisation factor.
The drawdown
\begin{equation}
\text{dd}_w(t) =
\frac{P_{t-1}}{\max_{j=1,\ldots,w} P_{t-j}} - 1,
\end{equation}
measures the distance of the current price from its recent peak,
where $P_t$ denotes the close price at bar $t$ and
$\max_{j=1,\ldots,w} P_{t-j}$ is the maximum close price over
the preceding $w$ bars.
The average true range normalised by the close price
\begin{equation}
\text{atr}_w(t) = \frac{1}{w\,P_{t-1}}
\sum_{j=1}^{w}\max\!\left(
H_{t-j} - L_{t-j},\;
|H_{t-j} - P_{t-j-1}|,\;
|L_{t-j} - P_{t-j-1}|
\right),
\end{equation}
measures price range volatility including gaps between consecutive
bars, where $H_t$ and $L_t$ denote the high and low prices at
bar $t$, $H_{t-j} - L_{t-j}$ is the intrabar range,
$|H_{t-j} - P_{t-j-1}|$ and $|L_{t-j} - P_{t-j-1}|$ measure
the distance of the high and low from the previous close,
and division by $P_{t-1}$ normalises the measure to be
scale-free.
The relative strength index (RSI)
\begin{equation}
\text{rsi}_w(t) = 100 -
\frac{100}{1 + \text{RS}_w(t)},
\qquad
\text{RS}_w(t) =
\frac{\frac{1}{w}\sum_{j=1}^{w}\max(r_{t-j},\,0)}
     {\frac{1}{w}\sum_{j=1}^{w}\max(-r_{t-j},\,0)},
\end{equation}
measures the relative magnitude of recent gains versus losses,
where $\text{RS}_w(t)$ is the ratio of the average positive
return to the average negative return over the past $w$ bars,
and $\text{rsi}_w(t) \in [0, 100]$ with values above 70
indicating overbought and below 30 indicating oversold conditions.

In addition,  we include four global features.
The high-low range normalised by the open price
\begin{equation}
\text{hl}(t) = \frac{H_{t-1} - L_{t-1}}{O_{t-1}},
\end{equation}
where $O_t$ denotes the open price at bar $t$, measures the
relative intrabar price range.
The volatility ratio
\begin{equation}
\text{vr}(t) = \frac{\text{std}_{10}(t)}{\text{std}_{120}(t)},
\end{equation}
compares short-term to long-term volatility, with values above
one indicating elevated short-term volatility relative to the
recent baseline.
The z-score
\begin{equation}
\text{zs}(t) = \frac{r_{t-1} - \bar{r}_{120,t}}{\text{std}_{120}(t)},
\end{equation}
measures how many standard deviations the most recent return
lies from its 120-bar rolling mean, where $\bar{r}_{120,t}$
and $\text{std}_{120}(t)$ are the rolling mean and standard
deviation over the preceding 120 bars.
The moving average deviation
\begin{equation}
\text{mad}(t) = \frac{P_{t-1} - \text{MA}_{20}(t)}{\text{MA}_{20}(t)},
\qquad
\text{MA}_{20}(t) = \frac{1}{20}\sum_{j=1}^{20} P_{t-j},
\end{equation}
measures the relative deviation of the current price from its
20-bar moving average, where $\text{MA}_{20}(t)$ is the simple
moving average of close prices over the preceding 20 bars.
This yields 22 price and volatility features in total.

For each of the six microstructure variables defined in
Section~\ref{sec:anomaly} -- effective spread, relative volume,
relative spread, Amihud illiquidity, Kyle's $\lambda$, and OIB -- we include both the value at bar $t-1$ and its
5-bar rolling mean
\begin{equation}
\bar{m}_5(t) = \frac{1}{5}\sum_{j=1}^{5} m_{t-j},
\end{equation}
where $m_{t-j}$ denotes the value of microstructure variable
$m$ at bar $t-j$. The rolling mean captures the persistence of
each liquidity condition over the preceding five bars,
yielding 12 microstructure features in total. The complete feature vector at bar $t$ is thus
$x_t \in \mathbb{R}^{34}$, comprising 22 price and volatility
features and 12 microstructure features, all computed from
information available strictly before bar $t$.

\begin{table}[!ht]
\centering
\scriptsize
\caption{AUC--ROC for one-bar-ahead anomaly prediction. The best result for each instrument--definition combination within each evaluation approach is shown in bold.}
\label{tab:aucroc}
\setlength{\tabcolsep}{5pt}
\begin{tabular}{llcccccccc}
\toprule
& & \multicolumn{4}{c}{Session-based} & \multicolumn{4}{c}{Cumulative} \\
\cmidrule(lr){3-6}\cmidrule(lr){7-10}
Instrument & Definition & LR & RF & XGB & LGBM & LR & RF & XGB & LGBM \\
\midrule
\multirow{5}{*}{VBTC.XE}
  & $D^{POT}$ & \textbf{0.812} & 0.810 & 0.782 & 0.781 & \textbf{0.821} & 0.815 & 0.793 & 0.792 \\
  & $D^{CV}$  & 0.729 & \textbf{0.731} & 0.685 & 0.685 & \textbf{0.732} & \textbf{0.732} & 0.690 & 0.688 \\
  & $D^{NR}$  & 0.720 & \textbf{0.723} & 0.693 & 0.694 & \textbf{0.732} & 0.730 & 0.704 & 0.705 \\
  & $D^{MR}$  & \textbf{0.755} & 0.741 & 0.695 & 0.697 & \textbf{0.762} & 0.746 & 0.703 & 0.704 \\
  & $D^{ANY}$ & \textbf{0.733} & 0.728 & 0.711 & 0.710 & \textbf{0.740} & 0.734 & 0.721 & 0.719 \\
\midrule
\multirow{5}{*}{VETH.XE}
  & $D^{POT}$ & 0.721 & \textbf{0.741} & 0.698 & 0.704 & 0.735 & \textbf{0.752} & 0.714 & 0.719 \\
  & $D^{CV}$  & 0.676 & \textbf{0.769} & 0.652 & 0.711 & 0.679 & \textbf{0.772} & 0.668 & 0.723 \\
  & $D^{NR}$  & \textbf{0.707} & 0.692 & 0.641 & 0.643 & \textbf{0.727} & 0.708 & 0.658 & 0.659 \\
  & $D^{MR}$  & \textbf{0.740} & 0.696 & 0.676 & 0.670 & \textbf{0.755} & 0.713 & 0.698 & 0.693 \\
  & $D^{ANY}$ & \textbf{0.722} & 0.714 & 0.665 & 0.685 & \textbf{0.739} & 0.727 & 0.686 & 0.704 \\
\midrule
\multirow{5}{*}{VIRBTC.ST}
  & $D^{POT}$ & \textbf{0.817} & 0.803 & 0.796 & 0.778 & \textbf{0.823} & 0.800 & 0.802 & 0.785 \\
  & $D^{CV}$  & 0.722 & \textbf{0.739} & 0.718 & 0.702 & 0.725 & \textbf{0.735} & 0.721 & 0.706 \\
  & $D^{NR}$  & 0.678 & \textbf{0.705} & 0.679 & 0.681 & 0.698 & \textbf{0.701} & 0.676 & 0.678 \\
  & $D^{MR}$  & \textbf{0.758} & \textbf{0.758} & 0.710 & 0.710 & \textbf{0.764} & 0.749 & 0.707 & 0.710 \\
  & $D^{ANY}$ & 0.716 & \textbf{0.739} & 0.708 & 0.716 & 0.724 & \textbf{0.732} & 0.708 & 0.714 \\
\midrule
\multirow{5}{*}{VIRSETHS.ST}
  & $D^{POT}$ & \textbf{0.701} & 0.695 & 0.678 & 0.672 & \textbf{0.708} & 0.702 & 0.681 & 0.676 \\
  & $D^{CV}$  & 0.664 & \textbf{0.755} & 0.707 & 0.699 & 0.700 & \textbf{0.774} & 0.722 & 0.716 \\
  & $D^{NR}$  & \textbf{0.678} & 0.635 & 0.585 & 0.577 & \textbf{0.696} & 0.652 & 0.587 & 0.579 \\
  & $D^{MR}$  & \textbf{0.719} & 0.659 & 0.558 & 0.567 & \textbf{0.698} & 0.666 & 0.579 & 0.579 \\
  & $D^{ANY}$ & \textbf{0.678} & \textbf{0.678} & 0.647 & 0.633 & \textbf{0.700} & 0.690 & 0.648 & 0.635 \\
\bottomrule
\end{tabular}
\end{table}

Table~\ref{tab:aucroc} presents the AUC--ROC values for one-bar-ahead anomaly prediction across four instruments, five anomaly targets, and four classifiers.
We report the results for both the session-based and cumulative feature approaches.
The $D^{POT}$ definition is the most predictable anomaly class. For VBTC.XE, LR and RF achieve values of 0.812 and 0.810 under the session-based approach. Under the cumulative approach, these values increase to 0.821 and 0.815. For VIRBTC.ST, LR achieves 0.817 and 0.823, respectively.
The cross-venue divergence definition $D^{CV}$ shows the most 
heterogeneous performance across instruments. For VETH.XE and 
VIRSETHS.ST, RF achieves AUC--ROC of 0.769 and 0.755 under 
the session-based scheme, rising to 0.772 and 0.774 under the cumulative 
approach. The good 
performance of RF on $D^{CV}$ suggests that the cross-venue 
pricing signal is captured more effectively by a non-linear classifier 
than by LR.

The $D^{NR}$ and $D^{MR}$ definitions show a moderate level of predictability. The AUC--ROC values range from 0.577 to 0.732 for $D^{NR}$ and from 0.558 to 0.764 for $D^{MR}$. For both definitions, LR is the best or joint-best classifier for most combinations of instruments and approaches.
The composite label $D^{ANY}$ gives AUC--ROC values between those of $D^{POT}$ and the less predictable definitions, ranging from 0.633 to 0.740. This is expected because the label combines four different types of events with different levels of predictability.

The cumulative approach gives higher AUC--ROC values in 68 of the 80 comparisons. The average improvement is approximately 0.009.
VIRSETHS.ST gives weaker results for most anomaly definitions, but performs well for $D^{CV}$. For $D^{NR}$, the best AUC--ROC values are 0.678 under the session-based approach and 0.696 under the cumulative approach. Both results are obtained by LR.

The precision and recall results reflect the strong class imbalance across the prediction targets, as anomaly events account for less than 1\% of valid bars. We also examined precision and recall at the test-set threshold that maximises the F1 score. However, since the thresholds were chosen using labels from the test set, these results are only descriptive and should not be interpreted as genuine out-of-sample performance.
 For this reason, they are not used to compare the models.
The F1-optimal thresholds range from 0.436 to 0.950, with XGBoost and LightGBM selecting higher values. Precision is low across all models and anomaly definitions and rarely exceeds 0.10, while recall varies more widely.
 LR and RF usually provide a better balance between precision and recall. On the other hand, the boosting models often achieve high recall but very low precision.
These results justify the use of AUC--ROC as the main evaluation metric.
 AUC--ROC evaluates how well a model separates the two classes across all classification thresholds, while precision and recall depend on the threshold selected.

\begin{table}[!ht]
\centering
\footnotesize
\caption{Most important features by permutation importance across instruments,
for each anomaly definition. Each cell shows the feature name and the number of instruments
(out of four) for which it ranks in the top 5.}
\label{tab:perm_summary}
\setlength{\tabcolsep}{5pt}
\begin{tabular}{lccc}
\toprule
Definition & Rank 1 & Rank 2 & Rank 3 \\
\midrule
$D^{POT}$ & \text{roll\_std\_15} (3) & \text{drawdown\_15} (3) & \text{drawdown\_5} (2) \\
$D^{CV}$  & \text{hl\_range} (3) & \text{atr\_5} (3) & \text{roll\_std\_15} (3) \\
$D^{NR}$  & \text{drawdown\_15} (3) & \text{vol\_ratio} (3) & \text{drawdown\_5} (3) \\
$D^{MR}$  & \text{rsi\_5} (4) & \text{vol\_ratio} (4) & \text{roll\_mean\_5} (4) \\
$D^{ANY}$ & \text{vol\_ratio} (4) & \text{roll\_std\_15} (3) & \text{atr\_5} (3) \\
\bottomrule
\end{tabular}
\end{table}

In addition to evaluating prediction performance, we examine which features contribute most to the model predictions. Table~\ref{tab:perm_summary} presents the features that are most important across instruments for each anomaly definition. The results are based on permutation importance calculated for the random forest classifier. A feature is included when it appears among the five most important features for a given instrument. The number in parentheses shows for how many of the four instruments this occurs.

The most important predictors vary across anomaly definitions. For $D^{POT}$, short-term rolling volatility and drawdown measures are the main predictors. The variables $\text{roll\_std\_15}$ and $\text{drawdown\_15}$ rank among the five most important features for three of the four instruments, while $\text{drawdown\_5}$ appears in the top five for two instruments. This result is consistent with the definition of $D^{POT}$, which captures extreme negative returns, and suggests that the classifier relies mainly on recent volatility and price declines.

For $D^{CV}$, the most important predictors are the high-low range (\text{hl\_range}), average true range (\text{atr\_5}), and rolling volatility (\text{roll\_std\_15}). Each appears among the five most important features for three instruments. This suggests that cross-venue pricing differences are more likely during periods of high intrabar price variation and greater market uncertainty.
For $D^{NR}$, the most important predictors are drawdown measures and the volatility ratio (\text{vol\_ratio}), which also appear in the top five for three instruments. The importance of drawdown measures suggests that these anomalies are often preceded by longer price declines rather than isolated short-term shocks.

The feature importance pattern for $D^{MR}$ shows the greatest similarity across the four instruments.
 The variables \text{rsi\_5}, \text{vol\_ratio}, and \text{roll\_mean\_5} appear among the five most important features for all four instruments. RSI and the rolling mean capture the positive short-term momentum required by this anomaly definition. The volatility ratio reflects the higher short-term volatility that often occurs around momentum reversals.

For the composite label $D^{ANY}$, the volatility ratio is the most consistent predictor and appears in the top five for all four instruments. Rolling volatility and ATR measures are also important. This is expected because $D^{ANY}$ combines four anomaly definitions that all capture unusual price behaviour in different ways.
Across all definitions, microstructure features are less important than price and volatility features. However, effective spread contributes to prediction for some instruments, especially VIRSETHS.ST, which has the lowest liquidity.

\section{Conclusions}
\label{sec:conclusions}

The European market for cryptocurrency ETPs has grown quickly in recent years and now attracts both individual and institutional investors. These products are also becoming a more common part of investment portfolios. It is important to understand how their prices change during the trading day and when unusual movements are more likely to occur. This information can help investors, risk managers, and regulators.
In this study, we analysed whether intraday anomalies in European cryptocurrency ETPs can be detected and predicted using high-frequency data. The analysis covers four Bitcoin and Ethereum ETPs traded on Xetra and Nasdaq Stockholm from January 2024 to December 2025. The data were aggregated into one-minute bars.

We considered four binary definitions of price anomalies. The baseline definition uses the POT method. It identifies one-minute bars in which the log-return is below a threshold estimated by fitting a GPD to left-tail observations.
The other three definitions complement the baseline approach by using cross-venue information or by examining price behaviour around extreme events. The cross-venue divergence anomaly identifies pricing differences between the two exchanges using the residual from an OLS regression of one venue’s returns on the returns of the other.
 The no-recovery anomaly identifies extreme price drops for which the cumulative log-return over the following ten bars remains below $30\%$ of the absolute initial log-return. It therefore measures limited recovery by the end of the observation window. Finally, the momentum-reversal anomaly identifies extreme price drops that follow a period of positive short-term momentum.

The analysis of the four anomaly definitions highlighted several important differences.
 The cross-venue divergence definition identifies a mostly separate group of events, as its pairwise Jaccard similarities with the other definitions are below 0.030. The no-recovery and momentum-reversal definitions show moderate overlap with each other and with the POT baseline.

Anomaly bars are generally associated with higher microstructure ratios, although some exceptions are observed for the cross-venue divergence definition. Instruments listed on Nasdaq Stockholm also show higher Amihud and Kyle's $\lambda$ ratios than the corresponding instruments listed on Xetra. These differences may be related to their lower general level of liquidity.
 Anomaly rates also change during the trading day and across the week. Unusual price movements occur more often near market opening and closing as well as on Fridays.

Using an out-of-sample prediction approach with four classifiers -- LR, RF, XGBoost, and LightGBM -- we showed that all four types of anomalies can be predicted one bar ahead. The AUC--ROC values range from 0.56 to 0.82. The POT definition is the most predictable class, with LR achieving AUC--ROC above 0.81 for the two Bitcoin instruments.
The cross-venue divergence definition is predicted most accurately by RF. This suggests that the information contained in cross-venue residuals may be better captured by non-linear models.
The no-recovery and momentum-reversal definitions show a moderate level of predictability. LR gives the best results for most combinations of instruments and anomaly definitions. The cumulative feature-conditioning approach also produces higher AUC--ROC values than the session-based approach. This suggests that information from earlier trading sessions may improve prediction.
The composite label $D^{ANY}$ is active when at least one anomaly definition is met. Its AUC--ROC values range from 0.63 to 0.74.

The permutation importance analysis showed that price and volatility features are generally more useful for predicting anomalies than microstructure variables. For $D^{POT}$, the main predictors are short-term rolling volatility and drawdown measures. Cross-venue divergence anomalies are predicted mainly by the high-low range, average true range, and short-term rolling volatility. For no-recovery anomalies, drawdown measures and the volatility ratio play the largest role. The main predictors of momentum-reversal anomalies are short-term RSI, the volatility ratio, and the rolling mean return, which rank among the five most important features for all four instruments. Microstructure variables generally play a smaller role, although their importance differs across instruments and is most pronounced for VIRSETHS.ST.
 Overall, the results show that the anomaly labels contain information that can be partially captured using recent intraday variables.

%%% Acknowledgements (if any)
%%% ------------------------------------------
\newpage
\section*{Acknowledgements}
We are grateful to Krzysztof Burnecki and Michał Balcerek for valuable remarks. 
%%% Supplementary materials (if any)
%%% ------------------------------------------

%%% Declaration of conflicting interests (should always be included)
%%% -----------------------------------------------------------------
\section*{Declaration of conflicting interests}
The authors declared no potential conflicts of interest with respect to the research, authorship and/or
publication of this article.
   %%% Alternatively, please, disclose here potential conflicting interests. 

%%% Funding (if any)
%%% ------------------------------------------
\section*{Funding}
The work of JK was supported by NCN Grant No. 2022/47/B/HS4/02139.

%%% Appendix (if any)
%%% ------------------------------------------
% \appendix
% \section*{Appendix}
% \section{Title of the first appendix section}
% This is the appendix text that authors want to include in the main paper not as supplementary materials.

% \section{Title of the second appendix section}

%%% References
%%% ------------------------------------------
\bibliographystyle{plainnat}
\bibliography{references}

\end{document}